\documentclass[sigconf]{acmart}

\AtBeginDocument{%
  }

\setcopyright{acmlicensed}
\copyrightyear{2026}
\acmYear{2026}
\acmDOI{XXXXXXX.XXXXXXX}
\acmConference[Submission Under Review]{Make sure to enter the correct
  conference title from your rights confirmation email}{--}{--}
\acmISBN{978-1-4503-XXXX-X/2018/06}
\renewcommand\footnotetextcopyrightpermission[1]{}
\usepackage{booktabs}
\usepackage{multirow}
\usepackage{array}
\usepackage{graphicx}
\usepackage[table]{xcolor}
\usepackage{booktabs}
\usepackage{microtype}
\usepackage[dvipsnames]{xcolor}
\usepackage{colortbl}
\usepackage{xcolor}
\definecolor{agentbg}{RGB}{237,239,254}

\usepackage{booktabs}
\usepackage{balance}
\usepackage{multirow}
\usepackage{arydshln}
\usepackage{makecell}
\usepackage{subcaption}
\usepackage[utf8]{inputenc} 
\usepackage{tabularx}
\usepackage{enumitem}
\usepackage[most]{tcolorbox}
\begin{document}

%%
%% The "title" command has an optional parameter,
%% allowing the author to define a "short title" to be used in page headers.
\title[\benchmarkname : A Benchmark for Precision Lung Cancer Diagnosis and Treatment]{\benchmarkname: Benchmarking Multimodal Real-World Clinical Reasoning for Precision Lung Cancer Diagnosis and Treatment}

%%
%% The "author" command and its associated commands are used to define
%% the authors and their affiliations.
%% Of note is the shared affiliation of the first two authors, and the
%% "authornote" and "authornotemark" commands
%% used to denote shared contribution to the research.
\author{Fangyu Hao}
\affiliation{%
  \institution{Beijing Univ. Posts \& Telecommun.}
  \country{China}
}
\email{haofangyu@bupt.edu.cn}

\author{Jiayu Yang}
\affiliation{%
  \institution{Beijing Univ. Posts \& Telecommun.}
  \country{China}
}
\email{jiayuyang@bupt.edu.cn}

\author{Yifan Zhu}
\authornote{Corresponding authors.}
\affiliation{%
  \institution{Beijing Univ. Posts \& Telecommun.}
  \country{China}
}
\email{yifan_zhu@bupt.edu.cn}

\author{Zijun Yu}
\author{Qicen Wu}
\author{Yunlong Wang}
\affiliation{%
  \institution{Beijing Univ. Posts \& Telecommun.}
  \country{China}
}
\email{flyfish@bupt.edu.cn}

\author{Jiawei Li}
\author{Yulin Liu}
\author{Xu Zeng}
\author{Guanting Chen}
\affiliation{%
  \institution{Beijing Univ. Posts \& Telecommun.}
  \country{China}
}
\email{cgt@bupt.edu.cn}

\author{Shihao Li}
\author{Zhonghong Ou}
\author{Meina Song}
\affiliation{%
  \institution{Beijing Univ. Posts \& Telecommun.}
  \country{China}
}
\email{zhonghong.ou@bupt.edu.cn}

\author{Mengyang Sun}
\affiliation{%
  \institution{Tsinghua Univ.}
  \country{China}
}
\email{sunmy19@mails.tsinghua.edu.cn}

\author{Haoran Luo}
\authornotemark[1] 
\affiliation{%
  \institution{Nanyang Technol. Univ.}
  \country{China}
}
\email{haoran.luo@ieee.org}

\author{Yu Shi}
\author{Yingyi Wang}
\affiliation{%
  \institution{Peking Union Med. Coll. Hosp.}
  \country{China}
}
\email{wangyingyi@pumch.cn}

%%
%% By default, the full list of authors will be used in the page
%% headers. Often, this list is too long, and will overlap
%% other information printed in the page headers. This command allows
%% the author to define a more concise list
%% of authors' names for this purpose.
\renewcommand{\shortauthors}{Hao et al.}
\newcommand{\benchmarkname}{LungCURE}
\newcommand{\agentname}{LCAgent}

%%
%% The abstract is a short summary of the work to be presented in the
%% article.
\begin{abstract}
Lung cancer clinical decision support demands precise reasoning across complex, multi-stage oncological workflows. Existing multimodal large language models (MLLMs) fail to handle guideline-constrained staging and treatment reasoning. 
We formalize three oncological precision treatment (OPT) tasks for lung cancer, spanning TNM staging, treatment recommendation, and end-to-end clinical decision support. We introduce \benchmarkname, the first standardized multimodal benchmark built from 1,000 real-world, clinician-labeled cases across more than 10 hospitals. 
We further propose \agentname, a multi-agent framework that ensures guideline-compliant lung cancer clinical decision-making by suppressing cascading reasoning errors across the clinical pathway. 
Experiments reveal large differences across various large language models (LLMs) in their capabilities for complex medical reasoning, when given precise treatment requirements. 
We further verify that \agentname, as a simple yet effective plugin, enhances the reasoning performance of LLMs in real-world medical scenarios. 
Project resources are available at https://joker-hfy.github.io/LungCURE/.
\end{abstract}



\maketitle

\section{Introduction}

\begin{figure}[]
  \setlength{\belowcaptionskip}{-0.5cm} 
  \centering
  %% width=\linewidth 在这里等于整个单栏的宽度
  \includegraphics[width=\linewidth]{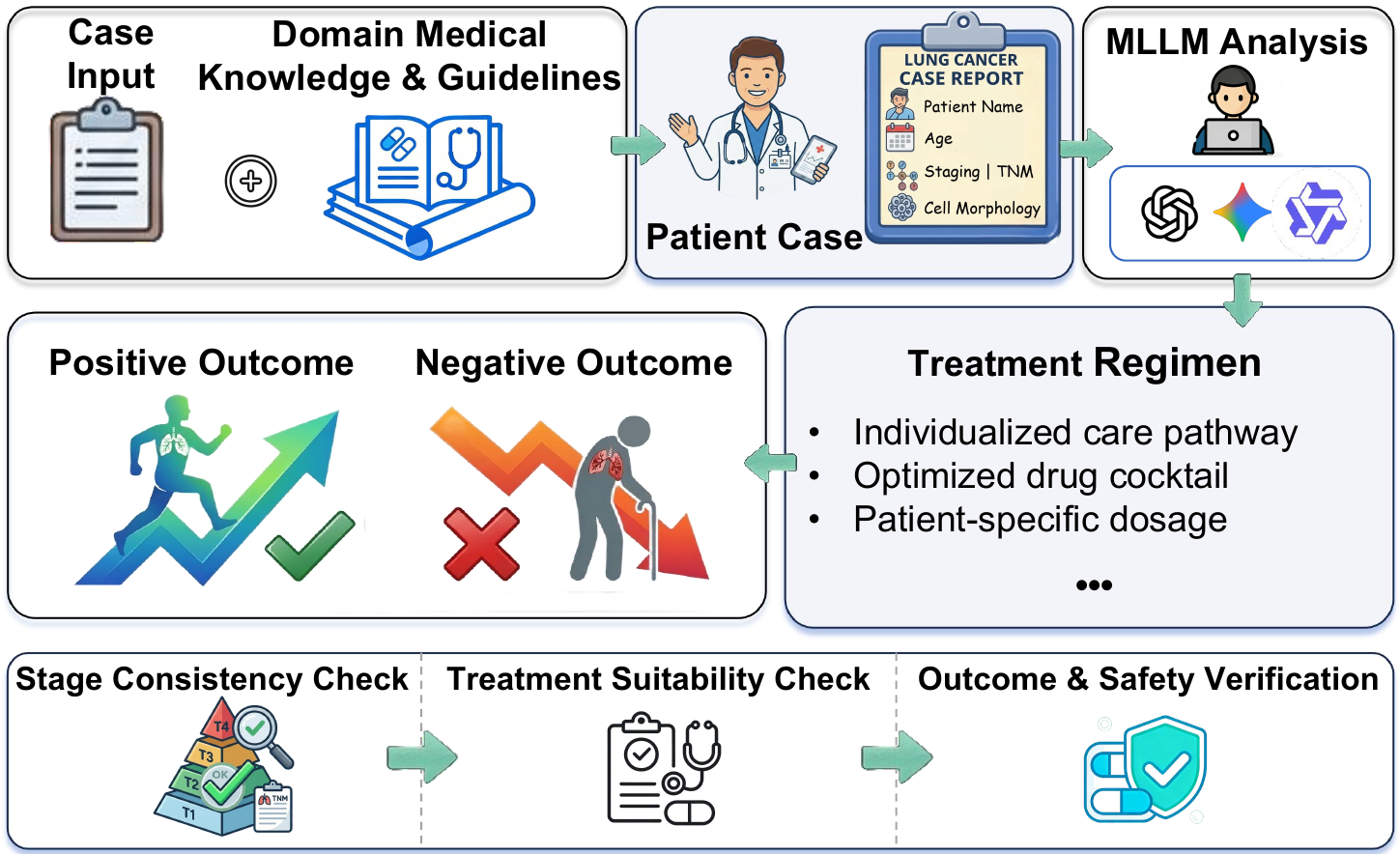}
  \caption{Task formulation of clinical treatment strategy generation driven by MLLMs}
  \label{fig:matrix}
\end{figure}

\begin{figure*}[t]
  \setlength{\belowcaptionskip}{-0.4cm} 
  \centering
  \includegraphics[width=\textwidth]{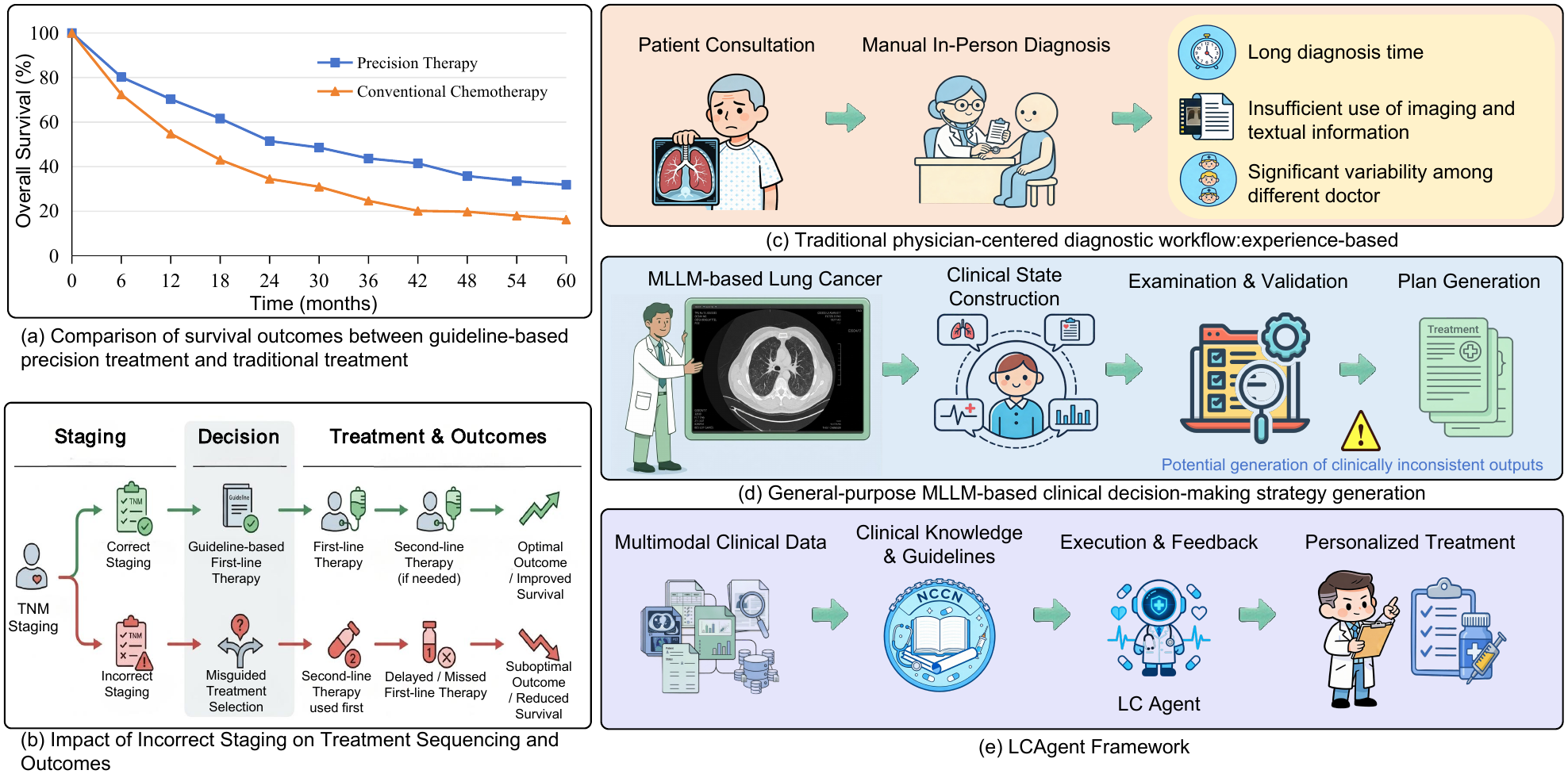} 
  \caption{Framework and Workflow of \agentname: From global mortality analysis to multimodal precision medicine}
  \label{fig:study_overview}
\end{figure*}

Lung cancer is considered one of the cancers with the highest incidence and mortality rates worldwide, and it is a key entry point for shifting from general chemotherapy to personalized oncological precision treatment (OPT) \cite{bray2024global}. 
Such precision treatment requires accurately determining the patient's current pathological stage \cite{lee2024lung} according to frequently updated medical guidelines (such as AJCC\footnote{\url{https://www.facs.org/}} , NCCN\footnote{\url{https://www.nccn.org/}} and CSCO\footnote{\url{https://www.csco.org.cn/}}), and deploying corresponding multi-line treatment regimens \cite{DBLP:journals/corr/abs-2509-07325}. 
The new multimodal large language models (MLLMs)-based paradigm \cite{li2023llava,DBLP:conf/emnlp/ChenGOGCCWCJWW24,zhu2024mmedpo} utilizing examination reports for lung cancer patients to perform staging assessment and treatment recommendations would significantly reduce the workload of clinicians, and therefore benefits patients in medically underdeveloped regions (Figure \ref{fig:matrix}) \cite{Rutunda2026}.
Unlike other cancer types \cite{DBLP:conf/miua/MeseguerAN25,lu2026gastricxmultimodalmultiphasebenchmark,guo2026mmneurooncomultimodalbenchmarkinstruction}, lung cancer involves approximately more than 100 staging combinations \cite{DETTERBECK2024882}, and deduces different treatment plans and prognoses based on driver genes and other clinical indicators \cite{captier2025integration}, posing a core challenge for MLLM-driven OPT.

%Precision treatment has been demonstrated to yield significant survival benefits compared to conventional therapeutic approaches(Figure \ref{fig:study_overview}(b)). Relative to physician doctor–driven decision-making (Figure \ref{fig:study_overview}(c)), MLLM-driven precision treatment offers superior decision consistency, positioning it as a highly promising paradigm for precision oncology.

Surprisingly, however, we observed that current mainstream MLLMs fail to adequately handle guideline-constrained staging and treatment reasoning\cite{DBLP:conf/emnlp/PanditXHWCXD25} required by precision therapy (Figure \ref{fig:study_overview}-a) \cite{DBLP:journals/corr/abs-2502-04381,Reck2021FiveYear}. 
The content they generate conversely deteriorates the quality of treatment \cite{xia2024cares}, potentially resulting in fatal outcomes (Figure \ref{fig:study_overview}-b) \cite{DBLP:journals/corr/abs-2403-14473}. Meanwhile, to the best of our knowledge, the differences in the capabilities of various MLLMs in assisting with lung cancer treatment decision-making remain to be quantitatively compared \cite{duan2025multi}.
Furthermore, traditional physician-centered workflows (Figure \ref{fig:study_overview}-c) also warrant comparison against MLLM-based approaches (Figure \ref{fig:study_overview}-d) \cite{DBLP:conf/ijcai/WuW024,DBLP:conf/naacl/BenkiraneKP25}.

Therefore, a critical research questions raises: 
\begin{tcolorbox}[notitle, rounded corners,colframe=gray, colback=gray!7, boxrule=1.5pt, boxsep=1.2pt, left=0.15cm, right=0.17cm, enhanced, shadow={1.5pt}{-1.5pt}{0pt}{opacity=5,gray!20},toprule=1.5pt, before skip=0.65em, after skip=0.75em]
\emph{
  {
    \centering 
  {
    \fontsize{8pt}{13.2pt}\selectfont 
    How can MLLMs be guided to generate clinically valid and guideline-compliant decisions for lung cancer, and how to quantitatively assess the capability of them?
  }
  \\
  }
  }
\end{tcolorbox}
To address this issue, we introduce a real-world Lung Cancer Benchmark for Clinical Understanding and Reasoning Evaluation, \benchmarkname,  centered on human doctors deriving consultated diagnoses and treatment plans from patient examination reports, and we perform a series of confirmatory explorations (Figure \ref{fig:study_overview}-e).
The main contributions are as follows:
\begin{itemize}[leftmargin=*, itemsep=0.3em, topsep=0pt]
\item \textbf{MLLM-driven OPT for Lung Cancer:} We formalize the research problem of MLLM-driven OPT for lung cancer, decomposing the lung cancer clinical decision support (CDS) workflow into three reasoning tasks: TNM staging, treatment recommendation, and end-to-end clinical decision support.

\item \textbf{\benchmarkname ~Benchmark:} We construct \benchmarkname, the first standardized multi-task multimodal benchmark for lung cancer clinical decision support, comprising 1,000 real-world clinical cases with expert-annotated gold standards.

\item \textbf{\agentname ~Framework:} We propose \agentname, a knowledge-guided multi-agent framework that boosts multiple state-of-the-art MLLMs in a plug-in way, and validate its effectiveness across \benchmarkname.
\end{itemize}

\begin{figure*}[htbp]
  \setlength{\belowcaptionskip}{-0.4cm} 
  \centering
  \includegraphics[width=0.9\textwidth]{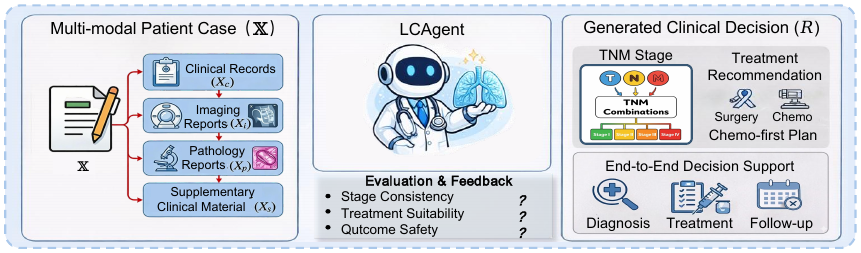} 
  \caption{The definition and overview of MLLM-driven OPT tasks for Lung Cancer.}
  \label{fig:definition}
\end{figure*}

\section{Related Work}
\textbf{Clinical Diagnosis and Decision Support.}
Clinical diagnosis and decision support systems (CDSS) have evolved across medicine from early rule-based systems encoding expert knowledge and clinical guidelines  \cite{susanto2023effects,DBLP:journals/jamia/SimGGHKLT01} to data-driven approaches leveraging large-scale electronic health records, imaging, laboratory tests, and genomic information \cite{DBLP:conf/nips/ChoiBSKSS16,DBLP:journals/jamia/XiaoCS18}. Early systems focused on standardizing decision-making and reducing inter-physician variability through structured protocols and decision trees \cite{sutton2020overview}. 
With the growing availability of heterogeneous medical data, recent work has increasingly applied machine learning and multimodal learning techniques to integrate diverse sources of information\cite{DBLP:conf/mm/ShiSS0YY24}, supporting tasks such as treatment recommendation and prognosis estimation \cite{wu2025harnessing,NEURIPS2024_90d1fc07,DBLP:conf/miccai/HuangZS25,Niu25}. These developments highlight the growing potential of CDSS to represent complex patient states and support multi-factor, multi-step clinical decision-making workflows across a broad range of diseases and specialties \cite{DBLP:conf/ijcai/UmerenkovNSARKK25}.

\textbf{Reasoning with Multimodal Large Language Models.}
Multimodal large language models (MLLMs) extend the capabilities of large language models to process and reason over diverse medical data modalities \cite{openai2024gpt4technicalreport,geminiteam2025geminifamilyhighlycapable,bai2025qwen3vltechnicalreport}. 
Building upon advances in natural language processing, vision-language pretraining, and instruction tuning \cite{DBLP:conf/mm/YangTS024,liu2023visual,dai2023instructblip}, MLLMs have been increasingly applied in the medical domain \cite{DBLP:conf/iclr/0005ZLWSWZ0Y25} for tasks such as automated report interpretation, information extraction, medical image understanding, and preliminary diagnostic reasoning \cite{DBLP:conf/mm/LiangZWZLW24,DBLP:journals/corr/abs-2407-02235,DBLP:conf/miccai/ShaPML25,DBLP:conf/miccai/XuSKW25}. 
Recent developments further explore their ability to perform multi-step and structured reasoning over heterogeneous patient data, model interdependent clinical variables \cite{xia2026mmedagentrloptimizingmultiagentcollaboration}, and generate coherent outputs that reflect complex clinical workflows \cite{DBLP:journals/corr/abs-2305-09617,kim2024llm}.

\section{MLLM-driven OPT Tasks for Lung Cancer}
As illustrated in Figure \ref{fig:definition}, we design three tasks, namely TNM staging, Treatment Recommendation and End-to-End Clinical Decision Support, aiming to simulate the real-world clinical workflow for lung cancer diagnosis and treatment.  
All the three tasks share the same patient multimodal input representation:

\begin{equation}
\mathbb{X} = \{X_m \mid m \in \mathcal{M}\}, \quad \mathcal{M} \subseteq \{C, I, P, S\},
\end{equation}
where $X_C$, $X_I$, $X_P$, and $X_S$ denote the clinical records, imaging reports, pathology reports, and supplementary clinical materials, respectively. Since not all modalities are available for every patient in real clinical scenarios, all three tasks allow missing modalities and require the model to perform reasoning under any available combination of inputs.

The three tasks differ in their input conditions and reasoning objectives, and are formally defined as follows:

\textbf{Task 1 (TNM Staging):} Given $\mathbb{X}$, predict the TNM staging result:
\begin{equation}
\hat{Y}_{\text{TNM}} = \arg\max_{Y} \Pr(Y \mid \mathbb{X}).
\end{equation}

\textbf{Task 2 (Treatment Recommendation):} Given $\mathbb{X}$ and the ground-truth of TNM stage $Y^*_{\text{TNM}}$, generate the conditioned treatment recommendation: 
\begin{equation}
\hat{R}_t = \arg\max_{R} \Pr(R \mid \mathbb{X}, Y^*_{\text{TNM}}).
\end{equation}

\textbf{Task 3 (End-to-End Decision Support):} Given $\mathbb{X}$ only, generate the clinical decision support recommendation without relying on any staging input:
\begin{equation}
\hat{R}_e = \arg\max_{R} \Pr(R \mid \mathbb{X}),
\end{equation}
where $Y_{\text{TNM}} \in \mathcal{Y}$ denotes the TNM staging label with $\mathcal{Y}$ being the set of all valid AJCC staging categories; $\hat{Y}_{\text{TNM}}$ denotes the model-predicted TNM stage; $Y^*_{\text{TNM}}$ denotes the expert-annotated ground-truth stage; $\hat{R}_t$ denotes the conditioned treatment recommendation generated with $Y^*_{\text{TNM}}$ as explicit input; and $\hat{R}_e$ denotes the unconditioned recommendation generated from $\mathbb{X}$ alone.

It should be noted that the key distinction between Task 2 and Task 3 lies in whether the ground-truth TNM stage is provided as a conditioning input. Task 3 is designed to reflect real-world clinical deployment, where a patient uploads 
multimodal clinical materials and receives an end-to-end decision support result 
without any manual staging intervention.
The performance gap between the two tasks is able to be used to quantitatively analyze how staging errors propagate into clinical decision making.

\begin{figure*}[t]
  \setlength{\belowcaptionskip}{-0.4cm} 
  \centering
  \includegraphics[width=0.9\textwidth]{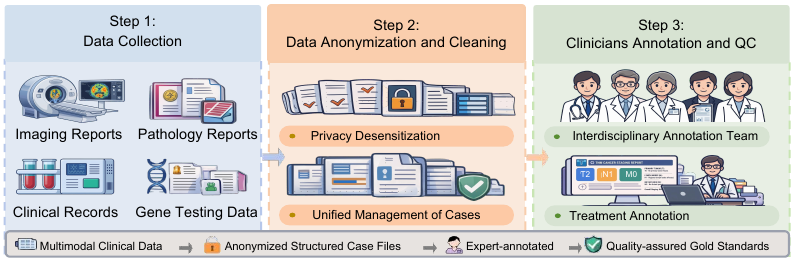} 
  \caption{Overview of the \benchmarkname~ construction pipeline.}
  \label{fig:definition and overview}
\end{figure*}

\section{The \benchmarkname ~Benchmark}

%In this section, we present a novel benchmark, namely \benchmarkname, specifically designed for training and evaluation of MLLM-driven Lung Cancer OPT Tasks. 

\subsection{Dataset Collection}
We construct \benchmarkname ~from 1,000 real-world clinical cases collected across more than ten hospitals in China between 2019 and 2025 (Figure \ref{fig:definition and overview}). 
The dataset comprises diverse multimodal clinical data, including imaging reports, pathology reports, clinical records, and genomic testing results. 
All data are fully de-identified to remove sensitive patient information and are systematically organized into unified, case-level documents that consolidate heterogeneous medical records into a standardized format. 
Data used in this study is from a retrospective study, and was approved by the Ethics Committee of Peking Union Medical College Hospital. All patients have signed an informed consent form before study enrollment.

To ensure high-qualitied ground truth, we adopt a two-stage expert annotation protocol, comprising evidence-based TNM staging with explicit reasoning and treatment plan generation based on structured clinical information by senior clinicians, forming reliable gold standards for evaluation. 
The dataset is now public available\footnote{\url{https://huggingface.co/datasets/Fine2378/LungCURE}}.
Further implementation details for each stage are provided in Appendix \ref{sec:lkbench}.

\subsection{Evaluation Metrics}
To systematically evaluate performance on the designed lung cancer OPT tasks, we define a set of evaluation metrics covering all the three tasks. 
The evaluation framework assesses not only the objective correctness of model outputs, but also the medical validity and clinical compliance of the reasoning process. 
For metrics involving subjective judgment, we adopt an LLM-as-a-Judge evaluation paradigm, in which a language model evaluator scores model outputs according to predefined rubrics (see Appendix \ref{sec:evaluation_prompts} for the specific evaluation prompts). 
Further details on these evaluation metrics are provided in Appendix \ref{sec:lkbench_metrics}.

\textbf{Evaluation on TNM Staging Task.}
The TNM staging evaluation measures the model's ability to infer tumor staging from multimodal clinical data, as well as the medical validity of its reasoning process. We establish two metrics for this task:
\begin{itemize}[leftmargin=*, itemsep=0.3em, topsep=0pt]
\item \textbf{TNM Staging Accuracy} evaluates the consistency between the model's predicted TNM stage and the ground-truth. 
Note that a prediction is considered correct if and only if the T, N, and M stages of the current case are all correct. 
\item \textbf{Reasoning Quality} evaluates the medical validity and evidence traceability of the model's reasoning process when generating TNM staging results by the judging model. 
\end{itemize}

\textbf{Evaluation on Clinical Decision Support Tasks.}
The clinical decision support (CDS) tasks evaluate the quality of treatment recommendations generated by the model. Both Task 2 and Task 3 are evaluated using the same three metrics: Precision and BERT-F1:
\begin{itemize}[leftmargin=*, itemsep=0.3em, topsep=0pt]
\item \textbf{Precision} measures the usefulness of the model-generated recommendation $\hat{R}$, refereed by the clinician's prescription $R^*$. 
\item \textbf{BERT-F1} measures the semantic similarity between the model's treatment decision reasoning process and the reference reasoning process provided by clinicians. 
\end{itemize}

\section{\agentname : A Simple Yet Effective Approach}
%\subsection{Overview of the Multi-Agent Architecture}
Existing general-purpose MLLMs exhibit systematic reasoning degradation and medical hallucination when applied to lung cancer clinical decision-making, failing to produce clinically valid and guideline-compliant diagnostic and therapeutic outputs. 
Thus, we propose \agentname, as a viable approach, which decomposes the lung cancer clinical decision-making workflow into two serially dependent stages with clearly delineated functional boundaries. 
By enforcing strict decision boundaries between stages and injecting expert prior knowledge at critical reasoning nodes, \agentname~ ensures logical consistency along the clinical pathway while effectively suppressing the accumulation and propagation of cascading reasoning errors.
Here we briefly introduce the method of \agentname, and a detailed formalization is presented in Appendix \ref{sec:method}. 
Our code is now public available \footnote{\url{https://github.com/Joker-hfy/LungCURE}}.
The detailed prompts for \agentname~ are provided in Appendix \ref{sec:agent_prompts}.

%\subsection{Anatomical Dimension Isolation for Decoupled TNM Staging}
\textbf{Anatomical Dimension Isolation for Decoupled TNM Staging.}
Existing end-to-end generation approaches are prone to cross- dimensional semantic interference when processing composite anatomical descriptions, leading to systematic errors in TNM stage assignment. To address this, we adopt an anatomical dimension decoupling strategy, fully isolating the evidence extraction and reasoning processes of the T, N, and M components into three concurrently executed specialized agents, whose outputs are subsequently aggregated by a deterministic rule-based node to produce the final staging conclusion, entirely eliminating the stochasticity introduced by free-form generation.

\begin{table*}[t]
\setlength{\abovecaptionskip}{-0.03cm} 
\centering
\caption{Results on TNM Staging, Treatment Recommendation, and End-to-End Decision Support.}
\label{tab:tnm_cdss_mllm_ocr}
\footnotesize
\setlength{\tabcolsep}{1pt}
\renewcommand{\arraystretch}{0.7}
\newcolumntype{Y}{>{\centering\arraybackslash}X}
\begin{tabularx}{\textwidth}{@{} l *{12}{Y} @{}}
\toprule
\multirow{2}{*}{\textbf{Models}}
& \multicolumn{4}{c}{\textbf{TNM Staging}}
& \multicolumn{4}{c}{\textbf{Treatment Recommendation}}
& \multicolumn{4}{c}{\textbf{End-to-End Decision Support}} \\
\cmidrule(lr){2-5} \cmidrule(lr){6-9} \cmidrule(lr){10-13}
& \multicolumn{2}{c}{\textbf{Acc(\%)}} & \multicolumn{2}{c}{\textbf{RQ}} & \multicolumn{2}{c}{\textbf{Precision(\%)}} & \multicolumn{2}{c}{\textbf{F1}} & \multicolumn{2}{c}{\textbf{Precision(\%)}} & \multicolumn{2}{c}{\textbf{F1}} \\
\cmidrule(lr){2-3} \cmidrule(lr){4-5} \cmidrule(lr){6-7} \cmidrule(lr){8-9} \cmidrule(lr){10-11} \cmidrule(lr){12-13}
& \textbf{ZH} & \textbf{EN} & \textbf{ZH} & \textbf{EN} & \textbf{ZH} & \textbf{EN} & \textbf{ZH} & \textbf{EN} & \textbf{ZH} & \textbf{EN} & \textbf{ZH} & \textbf{EN} \\
\midrule
\multicolumn{13}{c}{\textbf{\textit{MLLM (Image Input)}}} \\[-2pt]
\midrule
Kimi-K2.5        & 48.96 & 46.88 & 83.61 & 83.54 & 38.61 & 25.61 & 29.38 & 34.38 & 36.80 & 30.34 & 41.04 & 28.04 \\
Qwen3.5-397B     & \textbf{61.46} & \textbf{58.94} & \textbf{87.43} & \textbf{87.35} & 35.22 & 31.29 & \textbf{41.25} & 39.37 & 31.60 & 29.84 & 34.59 & 33.54 \\
GLM-4.6V         & 38.54 & 34.37 & 78.06 & 77.64 & 44.70 & 33.85 & 39.37 & \textbf{40.62} & \textbf{51.78} & 32.66 & \textbf{51.88} & 30.42\\
HuatuoGPT-Vision & 7.29  & 11.46 & 44.93 & 56.32 & -- & -- & -- & -- & -- & -- & -- & -- \\
DeepMedix-R1     & 0.00  & 1.04  & 26.32 & 26.94 & -- & -- & -- & -- & -- & -- & -- & -- \\
Llava-Med        & 0.00  & 0.00  & 21.46 & 20.56 & -- & -- & -- & -- & -- & -- & -- & -- \\
\noalign{\vspace{1pt}}
\cdashline{1-13}
\noalign{\vspace{1pt}}
Grok 4           & 1.04  & 11.51 & 58.26 & 64.36 & \textbf{65.48} & \textbf{40.04} & 34.38 & 40.00 & 47.75 & \textbf{33.07} & 31.25 & 35.02 \\
Claude Sonnet 4.6 & 25.00 & 28.13 & 78.19 & 80.69 & 25.39 & 22.99 & 38.13 & 30.00 & 32.08 & 27.62 & 37.39 & 25.41 \\
GPT-5.2          & 36.46 & 35.41 & 81.04 & 81.25 & 33.31 & 24.25 & 35.00 & 37.50 & 36.00 & 31.25 & 35.63 & 35.83 \\
Llama-4-maverick & 21.10 & 17.44 & 64.73 & 70.51 & 20.89 & 17.43 & 34.38 & 38.75 & 40.34 & 32.94 & 37.13 & \textbf{39.52} \\
\midrule
\multicolumn{13}{c}{\textbf{\textit{OCR + LLM (Text Input)}}} \\[-2pt]
\midrule
Kimi-K2.5        & 55.21 & 41.67 & 82.99 & \textbf{79.51} & 30.66 & 30.86 & 32.43 & 31.64 & 34.56 & 26.61 & 39.38 & 29.08 \\
Qwen3.5-397B     & \textbf{59.37} & 36.46 & \textbf{84.10} & 75.90 & 25.44 & 25.96 & \textbf{37.29} & 37.11 & 23.54 & 17.10 & 32.71 & 22.92 \\
GLM-4.6V         & 38.54 & 15.62 & 79.03 & 63.06 & \textbf{34.68} & \textbf{36.44} & 31.37 & \textbf{39.04} & 27.23 & \textbf{33.74} & 36.88 & 36.25 \\
HuatuoGPT-Vision & 6.81 & 11.21  & 42.15 & 51.69 & -- & -- & -- & -- & -- & -- & -- & -- \\
DeepMedix-R1     & 0.00 & 0.71  & 25.24 & 23.02 & -- & -- & -- & -- & -- & -- & -- & -- \\
Llava-Med        & 0.00 & 0.00 & 20.13 & 18.05 & -- & -- & -- & -- & -- & -- & -- & -- \\
\noalign{\vspace{1pt}}
\cdashline{1-13}
\noalign{\vspace{1pt}}
Grok 4           & 41.49 & \textbf{42.55} & 79.91 & 73.32 & 32.81 & 26.43 & 35.43 & 32.06 & \textbf{40.99} & 31.94 & \textbf{42.10} & 35.81 \\
Claude Sonnet 4.6 & 28.40 & 28.13 & 81.27 & 75.28 & 29.24 & 30.05 & 30.10 & 28.05 & 34.63 & 32.60 & 36.25 & 29.59 \\
GPT-5.2          & 38.54 & 31.25 & 79.30 & 73.68 & 19.94 & 25.93 & 33.27 & 36.90 & 35.13 & 29.90 & 40.00 & 35.42 \\
Llama-4-maverick & 22.92 & 10.48 & 77.15 & 59.35 & 23.83 & 11.80 & 33.91 & 36.84 & 31.00 & 32.93 & 38.96 & \textbf{37.92} \\
\bottomrule
\end{tabularx}
\end{table*}

% table 2
\begin{table*}[t]
\setlength{\abovecaptionskip}{-0.03cm} 
\setlength{\belowcaptionskip}{-0.2cm} 
\centering
\caption{Performance gains from \agentname~ across different base models.}
\label{tab:agent_enhancement_results}
\footnotesize
\setlength{\tabcolsep}{1pt}
\renewcommand{\arraystretch}{0.65}
\newcolumntype{Y}{>{\centering\arraybackslash}X}
\begin{tabularx}{\textwidth}{@{} l *{12}{Y} @{}}
\toprule
\multirow{2}{*}{\textbf{Models}}
& \multicolumn{4}{c}{\textbf{TNM Staging}}
& \multicolumn{4}{c}{\textbf{Treatment Recommendation}}
& \multicolumn{4}{c}{\textbf{End-to-End Decision Support}} \\
\cmidrule(lr){2-5} \cmidrule(lr){6-9} \cmidrule(lr){10-13}
& \multicolumn{2}{c}{\textbf{Acc(\%)}} & \multicolumn{2}{c}{\textbf{RQ}} & \multicolumn{2}{c}{\textbf{Precision(\%)}} & \multicolumn{2}{c}{\textbf{F1}} & \multicolumn{2}{c}{\textbf{Precision(\%)}} & \multicolumn{2}{c}{\textbf{F1}} \\
\cmidrule(lr){2-3} \cmidrule(lr){4-5} \cmidrule(lr){6-7} \cmidrule(lr){8-9} \cmidrule(lr){10-11} \cmidrule(lr){12-13}
& \textbf{ZH} & \textbf{EN} & \textbf{ZH} & \textbf{EN} & \textbf{ZH} & \textbf{EN} & \textbf{ZH} & \textbf{EN} & \textbf{ZH} & \textbf{EN} & \textbf{ZH} & \textbf{EN} \\
\midrule
\multicolumn{13}{c}{\textbf{\textit{MLLM (Image Input)}}} \\[-2pt]
\midrule
Qwen3.5-397B     & 61.46 & 58.94 & 87.43 & 87.35 & 35.22 & 31.29 & 41.25 & 39.37 & 31.60 & 29.84 & 34.59 & 33.54 \\

\rowcolor{agentbg}
\hspace{0.5em} + \agentname & 66.30 & 69.21 & 91.58 & 90.96 & 59.29 & 47.54 & 55.00 & 12.90 & 61.98 & 49.51 & 55.00 & 14.38 \\
Kimi-K2.5        & 48.96 & 46.88 & 83.61 & 83.54 & 38.61 & 25.61 & 29.38 & 34.38 & 36.80 & 30.34 & 41.04 & 28.04 \\
\rowcolor{agentbg}
\hspace{0.5em} + \agentname & 67.71 & 50.00 & 91.39 & 87.29 & 53.50 & 48.14 & 55.63 & 29.38 & 54.55 & 38.19 & 57.50 & 33.12 \\
GPT-5.2          & 36.46 & 35.41 & 81.04 & 81.25 & 33.31 & 24.25 & 35.00 & 37.50 & 36.00 & 31.25 & 35.63 & 35.83 \\
\rowcolor{agentbg}
\hspace{0.5em} + \agentname & 47.92 & 50.00 & 89.24 & 87.08 & 56.10 & 45.84 & 56.87 & 23.75 & 56.57 & 42.14 & 49.38 & 23.50 \\
\midrule
\multicolumn{13}{c}{\textbf{\textit{OCR + LLM (Text Input)}}} \\[-2pt]
\midrule
Qwen3.5-397B     & 59.37 & 36.46 & 84.10 & 75.90 & 25.44 & 25.96 & 37.29 & 37.11 & 23.54 & 17.10 & 32.71 & 22.92 \\
\rowcolor{agentbg}
\hspace{0.5em} + \agentname & 74.65 & 42.41 & 90.89 & 79.63 & 64.26 & 41.20 & 69.05 & 14.95 & 66.45 & 47.41 & 61.25 & 10.63 \\
Kimi-K2.5        & 55.21 & 41.67 & 82.99 & 79.51 & 30.66 & 30.86 & 32.43 & 31.64 & 34.56 & 26.61 & 39.38 & 29.08 \\
\rowcolor{agentbg}
\hspace{0.5em} + \agentname & 67.27 & 52.08 & 88.76 & 82.50 & 59.54 & 35.47 & 63.16 & 26.97 & 56.26 & 42.41 & 57.50 & 25.00  \\
GPT-5.2          & 38.54 & 31.25 & 79.30 & 73.68 & 19.94 & 25.93 & 33.27 & 36.90 & 35.13 & 29.90 & 40.00 & 35.42 \\
\rowcolor{agentbg}
\hspace{0.5em} + \agentname & 41.97 & 32.29 & 86.23 & 79.03 & 55.36 & 43.50 & 64.03 & 18.94 & 55.62 & 34.17 & 62.29 & 36.46 \\
\bottomrule
\end{tabularx}
\end{table*}

%\subsection{Feature Routing for Guideline-Grounded Treatment Recommendation}
\textbf{Feature Routing for Guideline-Grounded Treatment Recommendation.}
Building upon the deterministic staging output, the core challenge lies in the vast treatment decision state space of lung cancer, wherein injecting complete clinical guidelines into a single prompt induces severe attention dilution. 
To address this, we establish a deterministic scenario routing mechanism grounded in structured feature analysis. 
Critical decision variables are first extracted from the patient's multimodal records and standardized into a structured feature vector, which is subsequently mapped to the corresponding clinical scenario subspace. 
This mapping dynamically activates a scenario-specific expert agent that generates treatment recommendations under locally injected guideline subsets as hard constraints, ensuring all outputs are strictly grounded in evidence-based medicine.

\section{Experiments and Analysis} 
%\subsection{Overview}

To systematically evaluate the performance of multimodal large language models in lung cancer clinical workflows, we constructed \benchmarkname~ which comprises 1,000 real-world clinical cases. To enable efficient and controllable evaluation, we adopt a random sampling strategy to independently draw three subsets, forming the \benchmarkname-Core subset for primary experimental comparisons. Model performance is evaluated across all three task dimensions: TNM staging, treatment recommendation, and end-to-end clinical decision support. Table \ref{tab:tnm_cdss_mllm_ocr} reports the main experimental results on \benchmarkname-Core, while Table \ref{tab:agent_enhancement_results} presents the comparative performance of \agentname. ‘−’ denotes that the combined length of input and output exceeded the model's maximum supported sequence length, making evaluation infeasible. More experimental results can be found in Appendix \ref{sec:experiment}.

\begin{figure*}[t]
  \setlength{\belowcaptionskip}{-0.4cm} 
  \centering
  \includegraphics[width=\textwidth]{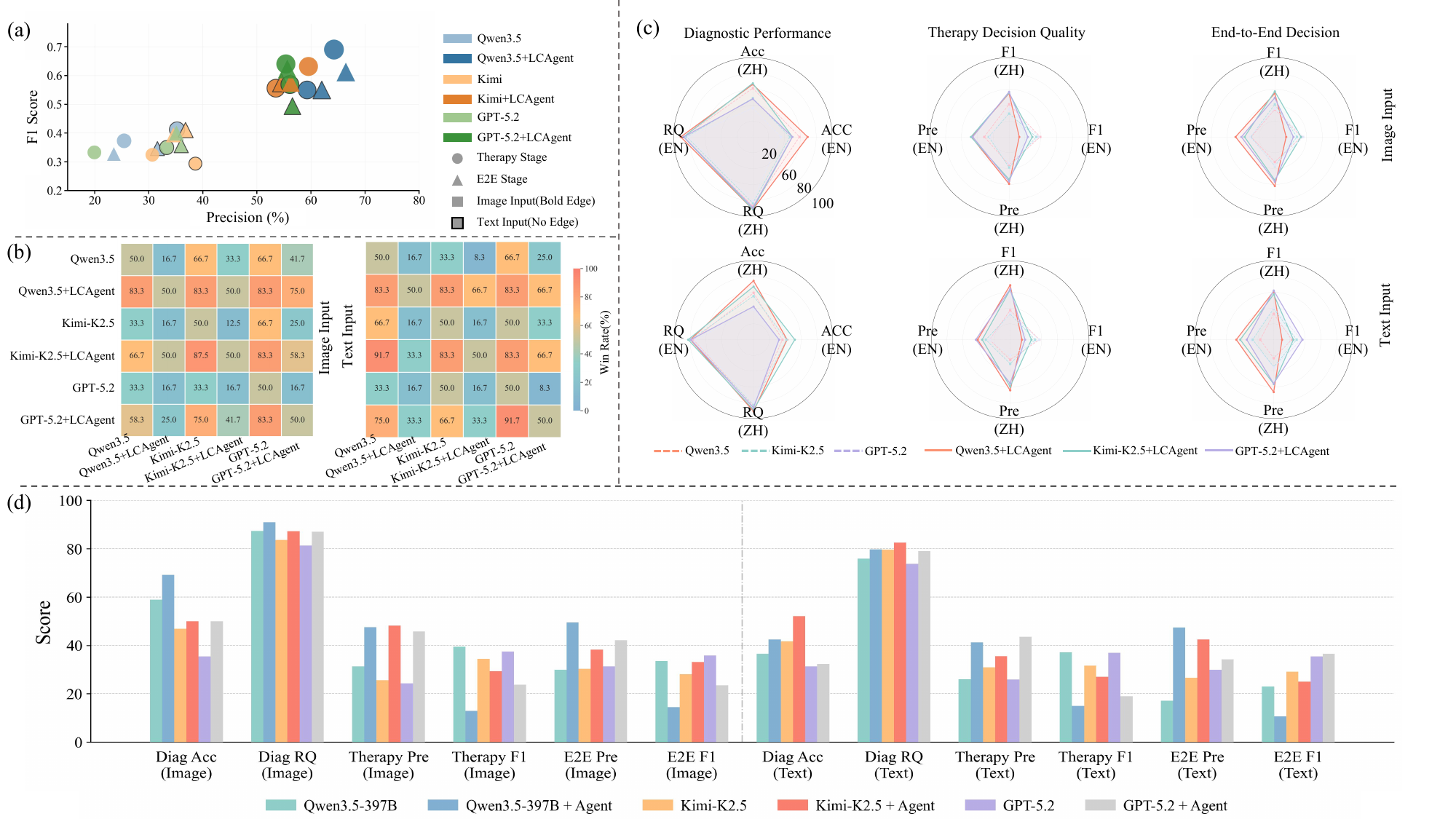} 
  \caption{Result Analysis. (a) F1-Precision Performance on \benchmarkname. (b) Pairwise win rate across VLLMs and LCAgent. (c) Performance evolution across clinical stages. (d) Overall comparison between VLLMs and \agentname.}
  \label{fig:F5}
\end{figure*}

%\subsection{Bench Effectively Evaluates and Differentiates MLLM Capabilities}
\textbf{Bench Effectively Evaluates and Differentiates MLLM Capabilities.}
Table \ref{tab:tnm_cdss_mllm_ocr} shows that the OPT diagnosis and treatment recommendation for lung cancer remains a highly challenging task for current MLLMs, and \benchmarkname~ effectively reveals fine-grained differences in model capabilities across clinical reasoning stages (Figure \ref{fig:F5}-a). In the TNM Staging task, even the best-performing model Qwen3.5 achieves an accuracy of only 61.46\% (ZH), while medical-specific models including HuatuoGPT, DeepMedix-R1, and Llava-Med perform at stochastic accuracy, indicating that domain specialization is not enough for precise structured clinical reasoning capability. Meanwhile, the benchmark also reveals clear performance stratification with inconsistent relative rankings across tasks. For instance, GLM-4.6V achieves the highest end-to-end F1 (51.88 ZH) despite unremarkable staging accuracy (38.54\%), suggesting that \benchmarkname~ decouples and independently assesses model capabilities at different clinical reasoning stages. Furthermore, most models exhibit systematic performance discrepancies between Chinese and English conditions (e.g., Grok~4 Treatment Recommendation precision: 65.48\% ZH vs.\ 40.04\% EN), and the OCR+LLM setting yields only marginal improvements over direct image input, confirming that the performance bottleneck primarily stems from clinical reasoning capability itself rather than input modality. These differentiated evaluation outcomes collectively validate \benchmarkname~ as an effective and discriminative benchmark for the lung cancer diagnosis and treatment task.

%\subsection{\agentname~ Performance}
\textbf{\agentname~ Performance.}
As observed in Table~\ref{tab:agent_enhancement_results}, the our proposed \agentname~ consistently and substantially outperforms the direct prompting baseline across almost all models, tasks, and input modalities. As further evidenced by the win-rate matrix, \agentname~ exhibits clear and consistent superiority over direct prompting baselines across all evaluated models (Figure~\ref{fig:F5}-b), further corroborating its comprehensive performance gains on the lung cancer clinical decision-making task. Taking Qwen3.5 under MLLM input as an example, \agentname~ improves end-to-end precision by 30.38\% and F1 by 59.01\%, while simultaneously improving Reasoning Quality from 87.43 to 91.58 (ZH), indicating that \agentname~ enhances not only the correctness of final decisions but also the quality of the underlying clinical reasoning process. The core mechanism underlying this improvement is that \agentname~ decomposes the complex clinical decision-making workflow into structured sub-stages with clearly defined responsibilities, enabling models to focus on a single reasoning objective at each stage, thereby effectively mitigating the pervasive evidence omission and cross-stage reasoning fragmentation under direct prompting.

Notably, the improvement margins of \agentname~ exhibit a meaningful differential distribution across tasks (Figure~\ref{fig:F5}-c): TNM Staging improvement is relatively moderate (+4.84\%), while Treatment Recommendation (+24.07\%) and End-to-End Decision Support (+30.38\%) show substantially larger gains. This pattern indicates that the primary benefit of \agentname~ derives from improving cross-stage information transmission and evidence integration. The TNM Staging task relies more heavily on the model's intrinsic medical knowledge and information extraction ability, leaving limited room for improvement through the Agent architecture. In contrast, Treatment Recommendation and End-to-End Decision Support involve multi-step reasoning and systematic construction of evidence chains, which are precisely the aspects where structured decomposition provides the greatest advantage. Under the OCR+LLM setting, performance improvements are even more pronounced: Qwen3.5-397B achieves\nolinebreak\ a +42.91\% increase in end-to-end precision and +38.82\% in Treatment Recommendation precision, both substantially exceeding the corresponding gains under the MLLM setting (Figure~\ref{fig:F5}-d). This observation can be attributed to the fact that in text-input scenarios, models become more dependent on systematic integration of structured textual information, making \agentname's structured decomposition even more critical for compensating this integration deficit. Furthermore, we also observe that \agentname's improvements are consistently stronger in Chinese than in English conditions, suggesting that the structured clinical reasoning workflow provides relatively greater benefit when processing Chinese medical records, likely due to the higher linguistic complexity and domain-specific terminology density in Chinese clinical documentation. Crucially, these improvements remain consistent across different backbone models including Kimi-K2.5 (end-to-end precision +17.75\%) and GPT-5.2 (end-to-end precision +20.57\%), confirming the strong model-agnostic generalizability of \agentname.
%Its effectiveness does not depend on specific model architectures or parameter scales, but rather originates from the systematic advantage inherent in the structured clinical reasoning workflow itself.

\section{Conclusion}
In this paper, we present \benchmarkname, the first standardized multimodal benchmark for real-world lung cancer clinical decision support, built from 1,000 real-world clinical cases across three tasks: TNM staging, treatment recommendation, and end-to-end decision support. 
Experiments reveal that current MLLMs exhibit persistent limitations in staging accuracy and cross-stage reasoning consistency. 
we also propose \agentname, a knowledge-guided multi-agent framework to show the further potential of measuring the knowledge-depended reasoning capacity of VLLMs.
% , demonstrating that reliable clinical decision support requires structured task decomposition and guideline-grounded reasoning rather than generic end-to-end generation alone.

%%
%% The acknowledgments section is defined using the "acks" environment
%% (and NOT an unnumbered section). This ensures the proper
%% identification of the section in the article metadata, and the
%% consistent spelling of the heading.

% \begin{acks}
% This work is supported by Beijing Municipal Natural Science Foundation under Grant L251042, the National Key Research and Development Program of China under Grant 2024YFC3308500,  National Natural Science Foundation of China under Grant 62406036, China Postdoctoral Science Foundation under Grant 2025M781457,  and also sponsored by the State Key Laboratory of Networking and Switching Technology under Grant NST20250110.
% \end{acks}

%%
%% The next two lines define the bibliography style to be used, and
%% the bibliography file.
\bibliographystyle{ACM-Reference-Format}
\bibliography{reference}

\clearpage

%%
%% If your work has an appendix, this is the place to put it.
\appendix

\section*{Appendix}
\setcounter{section}{0}

\section{\benchmarkname~ Construction Details}
\label{sec:lkbench}

\subsection{Dataset Construction Details}
\textbf{Step 1: Data Collection}\quad The \benchmarkname~ study data all come from real clinical cases in the Department of Medical Oncology at Peking Union Medical College Hospital. The study included lung cancer cases diagnosed by pathological examination between 2019 and 2025, strictly excluding cases with incomplete clinical data or unclear pathological diagnoses, ultimately including 1000 valid lung cancer cases. The included cases cover major pathological types of lung cancer, including adenocarcinoma, squamous cell carcinoma, and small cell lung cancer, encompassing the complete TNM stage range I-IV. For each case, multimodal diagnostic and treatment documentation was collected, stored in PDF or image format, including imaging reports, pathology reports, clinical records, and gene testing data. This also included structured clinical data such as patient basic information, tumor marker test results, TNM staging records, and clinical treatment plans, comprehensively matching the input requirements of TNM staging, CDSS, and end-to-end diagnostics.\\
\textbf{Step 2: Data Anonymization and Organization}\quad The privacy desensitization stage employs a comprehensive information masking strategy, thoroughly removing patient-related privacy information (name, ID number, hospital number, contact information, home address, etc.) from medical records. Simultaneously, sensitive content such as patient-provided external hospital reports and personal information related to treating physicians is uniformly deleted, eliminating all risks of privacy leaks. The case integration stage uses a single case as the sole index, structurally and uniformly integrating the scattered imaging reports, pathology reports, clinical records, and other PDF/image-format medical records for each lung cancer case into a single complete case PDF document. This achieves centralized collection and unified management of all medical information for a single case, providing a standardized and directly accessible data format for subsequent manual annotation and model evaluation.\\
\textbf{Step 3: Clinicians Annotation and Quality Control}
To construct a reliable gold standard for clinical decision-making, this study adopts a two-stage annotation protocol. In the TNM staging annotation phase, senior medical oncologists review the multimodal clinical documents of each case and systematically record the original evidential basis for each T, N, and M component. Uncertainty is explicitly annotated for evidence-insufficient findings, and an overall difficulty level is assigned to each case. Based on these annotations, the raw labels are further consolidated into a simplified gold standard that includes the final staging conclusions along with their corresponding reasoning evidence, serving as a reference benchmark for evaluating both the accuracy and reasoning quality of model-generated TNM staging. In the treatment annotation phase, standardized reference treatment plans are generated by clinical experts based on the annotated staging results and structured clinical information within each case—including histological subtype, driver gene status, PD-L1 expression level, and performance status—while strictly adhering to the NCCN and CSCO clinical guidelines. These treatment annotations serve as an objective benchmark for assessing the accuracy of model-generated therapeutic recommendations.

\subsection{Clinicians Annotation Protocol}
The construction of the \benchmarkname~ gold standard consists of two stages: TNM staging annotation and CDSS treatment plan generation, which differ methodologically. The former relies on expert-driven clinical judgment based on multimodal case documents, while the latter generates reference treatment plans strictly following clinical guidelines based on structured clinical information derived from expert annotations.
\subsubsection{TNM Staging Annotation}
TNM staging annotation is conducted a structured questionnaire by board-certified oncologists with expertise in thoracic malignancies, based on multimodal case documents (including imaging reports, pathology reports, laboratory tests, and genomic profiling results).

\textbf{T staging}: Annotators first assess whether the primary tumor is unassessable (Tx). If assessable, the T category (T1a–T4) is assigned based on maximum tumor diameter, and invasion characteristics are recorded, including visceral pleural invasion, central airway involvement, obstructive pneumonitis or atelectasis, invasion of adjacent structures (e.g., chest wall, diaphragm, mediastinum), major vascular invasion, and intrapulmonary metastases. Sites with insufficient evidence are marked as uncertain. When multiple T descriptors exist, the highest category is assigned following AJCC 8th edition. 

\textbf{N staging}: Annotators assess regional lymph node evaluability (Nx). If evaluable, the N stage (N0–N3) is assigned based on nodal involvement, and each involved station is recorded sequentially (ipsilateral peribronchial, hilar, mediastinal; subcarinal; contralateral mediastinal, hilar; supraclavicular). Suspicious but unconfirmed nodes are noted as uncertain. The highest N category is selected if multiple levels are involved. 

\textbf{M staging}: Annotators first determine M0 status. In cases of distant metastasis, M stage is categorized as M1a (contralateral lung or pleural/pericardial), M1b (single extrathoracic metastasis), or M1c (multiple metastases). Each metastatic site (bone, brain, liver, adrenal, or non-regional lymph nodes) is documented. Radiographically suspicious but unconfirmed lesions are recorded in an uncertainty field. Multiple metastatic features default to the highest M category. 

\textbf{Generation of Simplified Ground Truth}:
From raw structured annotations, a simplified ground truth is generated for each case, summarizing final T/N/M stages along with diagnostic reasoning, supporting automated quality assessment for model inference. 

\subsubsection{Treatment Plan Generation}
Reference treatment plans are generated by senior clinicians based on structured clinical variables derived (e.g., TNM stage, histology, driver mutations, PD-L1 expression, and treatment history), following NCCN and CSCO guidelines. Guideline discrepancies and missing critical information are explicitly documented. The final plans include treatment strategies, core drug regimens, and key considerations, serving as a standardized benchmark for evaluating model-generated treatment recommendations. 
\subsubsection{Quality Control}
Upon completion of TNM staging annotations, all entries are reviewed by independent quality control personnel. The review focuses on completeness, consistency between uncertainty annotations and supporting clinical evidence, and alignment between the reasoning evidence in the simplified gold standard and the original annotations. Any ambiguous or questionable entries are returned to the original annotators for verification before inclusion in the final dataset.

\section{Evaluation Metrics Details}
\label{sec:lkbench_metrics}
\subsection{TNM Staging Task Evaluation}

\begin{itemize}[leftmargin=*, itemsep=0.3em, topsep=0pt]
\item \textbf{TNM Staging Accuracy.} For each sample, the predicted TNM stage is directly compared against the expert annotation; a prediction is considered correct only if all components match exactly. The overall accuracy is reported at the dataset level.

\item \textbf{Reasoning Quality.} The evaluator scores four components separately---T stage, N stage, M stage, and overall synthesis---each on a scale of 1 to 5, and the final score is the average across all components. The scoring criteria focus on: (1) whether evidence is accurately traced to the source; (2) whether the reasoning for each individual stage component establishes sound clinical logic; and (3) whether the synthesis adheres to standard oncological staging rules.
\end{itemize}

\subsection{Clinical Decision Support Task Evaluation}

\begin{itemize}[leftmargin=*, itemsep=0.3em, topsep=0pt]
\item \textbf{Precision.} The evaluator compares the model output against the reference across treatment strategy, key medications, and clinical pathway, and computes the overall degree of alignment.

\item \textbf{BERT-F1.} This metric effectively reflects the degree to which the model's clinical decision-making reasoning aligns with expert clinical thinking, and is computed independently for Task~2 and Task~3.
\end{itemize}

\section{Methodology}
\label{sec:method}
\subsection{Multi-Agent Architecture}

Formally, the clinical decision-making task aims to find the optimal treatment strategy $\mathcal{T}^*$ given a patient's multi-modal medical record $\mathcal{R}$ and a vast set of clinical guidelines $\mathcal{G}$. This can be formulated as a conditional probability maximization problem:
\begin{equation}
    \mathcal{T}^* = \arg\max_{\mathcal{T}} \text{Pr}(\mathcal{T} \mid \mathcal{R}, \mathcal{G})
    \label{eq:prob_max}
\end{equation}

Direct generation approaches mapping $\mathcal{R}$ to $\mathcal{T}$ via a single monolithic prompt typically fail to align with stringent clinical guidelines. To address this deficiency, we formulate the lung cancer clinical decision-making process as a rule-constrained, step-by-step reasoning problem. We propose a \agentname~ framework. Our framework formulates the clinical workflow as a Directed Acyclic Graph (DAG) of functions, systematically decomposing the joint probability into two deterministic stages: 
\begin{equation}
    \mathcal{T}^* = \Psi_{\text{CDSS}} \Big( \Phi_{\text{stage}}\big(\mathcal{M}_{\text{percept}}(\mathcal{R})\big), \mathcal{G} \Big)
    \label{eq:dag_pipeline}
\end{equation}
where $\mathcal{M}_{\text{percept}}(\cdot)$ represents the neural perception agents responsible for semantic extraction, $\Phi_{\text{stage}}(\cdot)$ denotes the symbolic algorithmic logic gates for TNM staging, and $\Psi_{\text{CDSS}}(\cdot)$ is the scenario-specific expert routing mechanism for treatment recommendation. By establishing strict decision boundaries and injecting expert prior knowledge at specific nodes, this framework ensures consistent logical fidelity to clinical pathways and effectively mitigates cascading reasoning errors.

\subsection{Anatomical Dimension Isolation for Decoupled TNM Staging}

To resolve the compound spatial errors inherent in TNM staging, we introduce an anatomically-decoupled TNM staging pipeline that isolates the evidence extraction and reasoning of each T, N, and M component into dedicated agents, proceeding as follows: 

\paragraph{1) Semantic Standardization and Feature Routing:} We first employ a document-extraction agent $\mathcal{M}_{\text{extract}}$ to parse unstructured multi-modal reports $\mathcal{R}$ (e.g., CT, PET/CT, pathology reports). To prevent spatial logic confusion, we introduce a \textit{Composite Anatomical Site Splitting} algorithm. For instance, composite phrases like ``bilateral hilar and mediastinal nodes'' are forced to split into independent entities. This algorithm projects the raw text into three decoupled anatomical feature sets for Tumor ($E_T$), Node ($E_N$), and Metastasis ($E_M$):
\begin{equation}
    \{E_T, E_N, E_M\} = \mathcal{M}_{\text{extract}}(\mathcal{R} \mid \pi_{\text{extract}})
    \label{eq:feature_routing}
\end{equation}
where $\pi_{\text{extract}}$ is the prompt enforcing baseline laterality anchoring (e.g., distinguishing ipsilateral from contralateral lesions based on the primary tumor).

\paragraph{2) Independent Staging Agents:}  Three specialized LLM agents ($\mathcal{M}_T$, $\mathcal{M}_N$, and $\mathcal{M}_M$) process their respective feature sets concurrently. Each agent acts under rigorous Rule-Constrained Chain-of-Thought (RC-CoT) $\pi_{k}$. For example, the T-Agent strictly executes an absolute maximum diameter extraction rule, while the M-Agent evaluates distant metastasis via multi-organ combinatorial logic. The generation process is formalized as:
\begin{equation}
    s_k, u_k = \mathcal{M}_{k}(E_k \mid \pi_k), \quad \forall k \in \{T, N, M\}
    \label{eq:independent_agents}
\end{equation}
where $s_k$ represents the deterministic sub-stage (e.g., $T2a$), and $u_k$ represents the set of ``uncertain/suspicious'' nodes (e.g., ``nature to be determined'') identified during reasoning.

\paragraph{3) Deterministic Aggregation and Uncertainty Projection:} Finally, the independent outputs are aggregated using a deterministic code execution node $\Gamma_{\text{AJCC}}(\cdot)$. Rather than generating the final stage via LLM, this node utilizes a strict logic matrix derived from the AJCC manual to compute the comprehensive stage $S_{\text{final}}$ (e.g., IA1, IIIA):
\begin{equation}
    S_{\text{final}} = \Gamma_{\text{AJCC}}(s_T, s_N, s_M)
    \label{eq:aggregation}
\end{equation}
Furthermore, we propose a novel \textit{Uncertainty Projection Mechanism} $\Omega(\cdot)$ that calculates potential stage shifts caused by uncertain features $\mathcal{U} = u_T \cup u_N \cup u_M$. This mechanism yields a set of potential stages $\mathbb{S}_{\text{potential}} = \Omega(\mathcal{U}, S_{\text{final}})$, providing oncologists with actionable diagnostic alerts regarding how subsequent biopsies might alter the clinical stage.

\begin{table*}[t]
\setlength{\abovecaptionskip}{-0.03cm} 
\centering
\caption{Results on TNM Staging.}
\label{tab:tnm}
\footnotesize
\setlength{\tabcolsep}{1pt}
\renewcommand{\arraystretch}{0.7}
\newcolumntype{Y}{>{\centering\arraybackslash}X}
\begin{tabularx}{\textwidth}{@{} l *{12}{Y} @{}}
\toprule
\multirow{2}{*}{\textbf{Models}}
& \multicolumn{4}{c}{\textbf{T Staging}}
& \multicolumn{4}{c}{\textbf{N Staging}}
& \multicolumn{4}{c}{\textbf{M Staging}} \\
\cmidrule(lr){2-5} \cmidrule(lr){6-9} \cmidrule(lr){10-13}
& \multicolumn{2}{c}{\textbf{Acc(\%)}} & \multicolumn{2}{c}{\textbf{RQ}} & \multicolumn{2}{c}{\textbf{Acc(\%)}} & \multicolumn{2}{c}{\textbf{RQ}} & \multicolumn{2}{c}{\textbf{Acc(\%)}} & \multicolumn{2}{c}{\textbf{RQ}} \\
\cmidrule(lr){2-3} \cmidrule(lr){4-5} \cmidrule(lr){6-7} \cmidrule(lr){8-9} \cmidrule(lr){10-11} \cmidrule(lr){12-13}
& \textbf{ZH} & \textbf{EN} & \textbf{ZH} & \textbf{EN} & \textbf{ZH} & \textbf{EN} & \textbf{ZH} & \textbf{EN} & \textbf{ZH} & \textbf{EN} & \textbf{ZH} & \textbf{EN} \\
\midrule
\multicolumn{13}{c}{\textbf{\textit{MLLM (Image Input)}}} \\[-2pt]
\midrule
Kimi-K2.5        & 62.50 & 57.29 & 77.50 & 73.12 & 78.12 & 81.25 & 88.54 & 87.29 & 79.16 & 84.38 & 84.79 & 90.21 \\
Qwen3.5-397B     & 69.79 & 74.70 & 80.00 & 82.71 & 84.38 & 83.17 & 90.21 & 89.46 & 91.67 & 84.18 & 92.08 & 89.89 \\
GLM-4.6V         & 56.25 & 48.96 & 73.96 & 69.38 & 71.88 & 76.04 & 84.79 & 86.67 & 66.66 & 66.67 & 75.42 & 76.88 \\
HuatuoGPT-Vision & 16.67  & 25.00 & 41.88 & 50.21 & 29.17 & 29.17 & 49.58 & 52.92 & 29.17 & 59.38 & 43.33 & 65.83 \\
DeepMedix-R1     & 6.25  & 3.13  & 26.67 & 25.62 & 4.17 & 5.21 & 25.62 & 26.46 & 10.42 & 10.42 & 26.67 & 28.75 \\
Llava-Med        & 3.13  & 3.13  & 20.42 & 20.83 & 4.16 & 4.16 & 22.08 & 20.83 & 0.00 & 0.00 & 21.87 & 20.00 \\
\noalign{\vspace{1pt}}
\cdashline{1-13}
\noalign{\vspace{1pt}}
Grok 4           & 20.83 & 27.18 & 55.83 & 52.68 & 19.79 & 51.13 & 55.83 & 68.02 & 52.08 & 60.91 & 63.12 & 72.37 \\
Claude Sonnet 4.6 & 35.42 & 37.50 & 66.46 & 67.71 & 67.71 & 71.87 & 85.42 & 87.08 & 72.92 & 79.16 & 82.71 & 87.29 \\
GPT-5.2          & 53.13 & 50.00 & 74.17 & 70.42 & 66.67 & 69.79 & 83.54 & 86.04 & 77.09 & 80.21 & 85.42 & 87.29 \\
Llama-4-maverick & 31.45 & 27.88 & 58.58 & 56.25 & 38.69 & 58.19 & 66.54 & 75.61 & 55.59 & 73.07 & 69.07 & 79.66 \\
\midrule
\multicolumn{13}{c}{\textbf{\textit{OCR + LLM (Text Input)}}} \\[-2pt]
\midrule
Kimi-K2.5        & 71.87& 62.50 & 74.17 & 71.46 & 76.04 & 67.71 & 86.46 & 79.79 & 85.42 & 80.21 & 88.33 & 87.29 \\
Qwen3.5-397B     & 69.79 & 55.21 & 71.87 & 65.21 & 85.42 & 65.63 & 90.83 & 76.46 & 87.50 & 83.33 & 89.58 & 86.04 \\
GLM-4.6V         & 55.21 & 33.33 & 65.62 & 57.29 & 76.04 & 44.79 & 86.87 & 65.21 & 78.13 & 55.21 & 84.58 & 66.67 \\
HuatuoGPT-Vision & 14.58 & 22.92 & 50.83 & 46.87 & 52.08 & 25.00 & 70.42 & 50.83 & 59.37 & 66.67 & 68.54 & 68.54 \\
DeepMedix-R1     & 0.00 & 6.25  & 41.04 & 36.67 & 20.83 & 0.00 & 36.67 & 21.88 & 28.12 & 7.29 & 40.83 & 24.17 \\
Llava-Med        & 3.13 & 4.17 & 20.42 & 22.09 & 4.16 & 4.16 & 21.25 & 21.87 & 0.00 & 3.13 & 20.21 & 22.29 \\
\noalign{\vspace{1pt}}
\cdashline{1-13}
\noalign{\vspace{1pt}}
Grok 4           & 53.91 & 53.98 & 68.97 & 61.93 & 75.16 & 65.37 & 86.00 & 73.94 & 78.75 & 77.82 & 84.77 & 84.08 \\
Claude Sonnet 4.6 & 39.99 & 41.66 & 66.12 & 62.71 & 77.82 & 60.42 & 89.22 & 79.79 & 82.19 & 79.17 & 88.48 & 83.33 \\
GPT-5.2          & 54.17 & 54.17 & 67.29 & 66.46 & 66.67 & 54.17 & 84.37 & 75.21 & 78.13 & 75.00 & 86.25 & 79.37 \\
Llama-4-maverick & 34.37 & 23.19 & 60.62 & 51.83 & 68.75 & 34.91 & 83.33 & 58.80 & 85.42 & 65.36 & 87.50 & 67.43 \\
\bottomrule
\end{tabularx}
\end{table*}

\subsection{Feature Routing for Guideline-Grounded Treatment Recommendation}

Building upon the precise TNM stage $S_{\text{final}}$, we advance the pipeline to therapeutic decision-making:

\paragraph{1) Structured Profiling \& Algorithmic Triage:} A specialized agent first extracts critical decision-making factors (e.g., Histology, PS Score, PD-L1 expression) from the patient record, standardizing them into a profile vector $V_{\text{profile}}$. A deterministic routing script $\Phi_{\text{route}}$ acts as a clinical triage system, mapping the combination of $V_{\text{profile}}$ and $S_{\text{final}}$ to a specific clinical scenario subspace $\mathcal{C}_{id}$ (e.g., ``Early-Stage Post-Radical Resection''):
\begin{equation}
    \mathcal{C}_{id} = \Phi_{\text{route}}(V_{\text{profile}}, S_{\text{final}})
    \label{eq:routing}
\end{equation}

\paragraph{2) In-Context Guideline Injection and Recommendation:} Based on the triage result $\mathcal{C}_{id}$, the system dynamically activates a corresponding Expert Agent $\mathcal{M}_{\text{expert}}$. Highly dense, localized clinical guidelines and landmark trial literatures $\mathcal{G}_{id} \subset \mathcal{G}$ (e.g., NCCN/CSCO protocols, KEYNOTE series) are retrieved and injected as hard constraints. The final treatment recommendation $\mathcal{T}_{\text{final}}$ is generated as:
\begin{equation}
    \mathcal{T}_{\text{final}} = \mathcal{M}_{\text{expert}}\big(\mathcal{T} \mid \pi_{\text{expert}}(\mathcal{C}_{id}), \mathcal{G}_{id}, V_{\text{profile}}\big)
    \label{eq:expert_generation}
\end{equation}
To ensure clinical safety, we implement a \textit{Missing Value Handling Constraint}: if critical components of $V_{\text{profile}}$ are null, $\mathcal{M}_{\text{expert}}$ is forced to issue a pre-emptive clinical evaluation warning before attempting any recommendation. This explicit routing drastically optimizes token usage and ensures the outputs are fully grounded in Evidence-Based Medicine (EBM).
\section{Evaluation Prompts for \benchmarkname}
\label{sec:evaluation_prompts}
This section provides the detailed prompts used by the LLM judge for evaluation in the \benchmarkname. Due to formatting constraints, the prompts are presented in Figures 6–8.
\begin{figure*}[t]
  \centering
  \includegraphics[width=\textwidth]{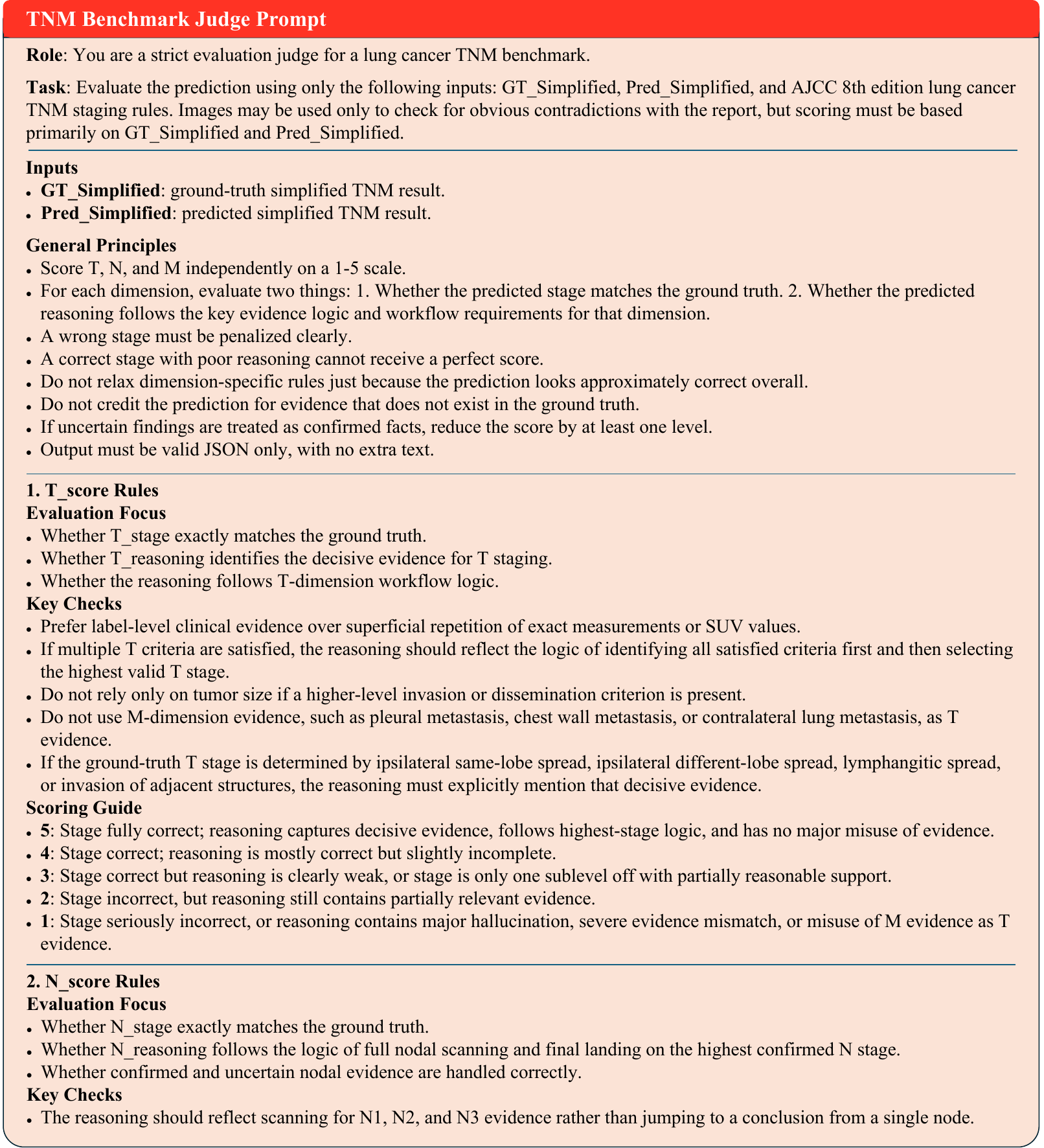} 
\end{figure*}

\begin{figure*}[t]
  \centering
  \includegraphics[width=\textwidth]{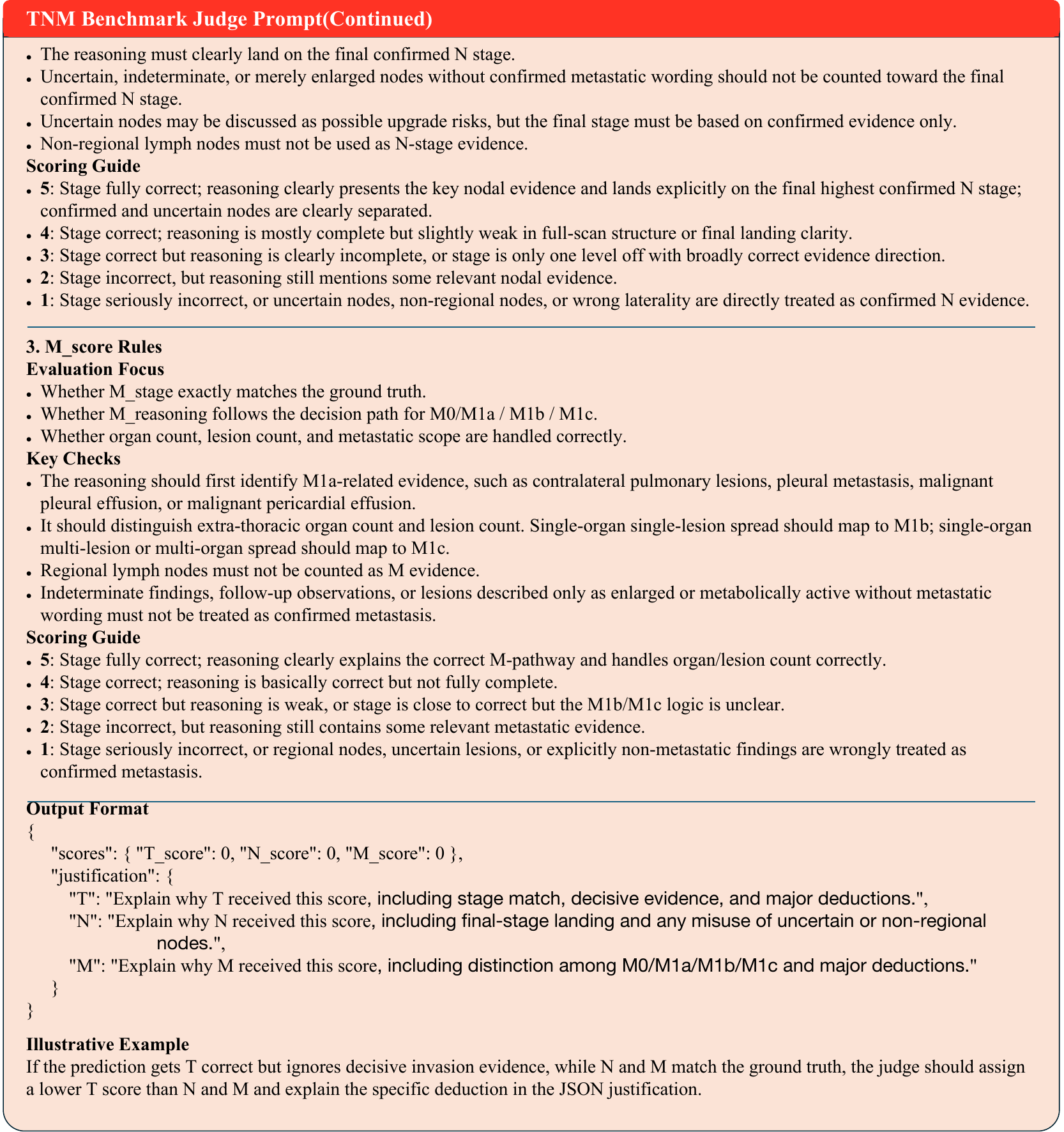} 
  \caption{Prompt used by the TNM Benchmark Judge. This prompt instructs the LLM to act as a strict evaluation judge for a lung cancer TNM benchmark, scoring predictions on a 1-5 scale based on stage accuracy and reasoning logic. It enforces a rigorous deduction policy, penalizing hallucinations, incorrect stages, or the misuse of evidence dimensions, and requires a structured JSON justification for the assigned scores.}
\end{figure*}

\begin{figure*}[t]
  \centering
  \includegraphics[width=\textwidth]{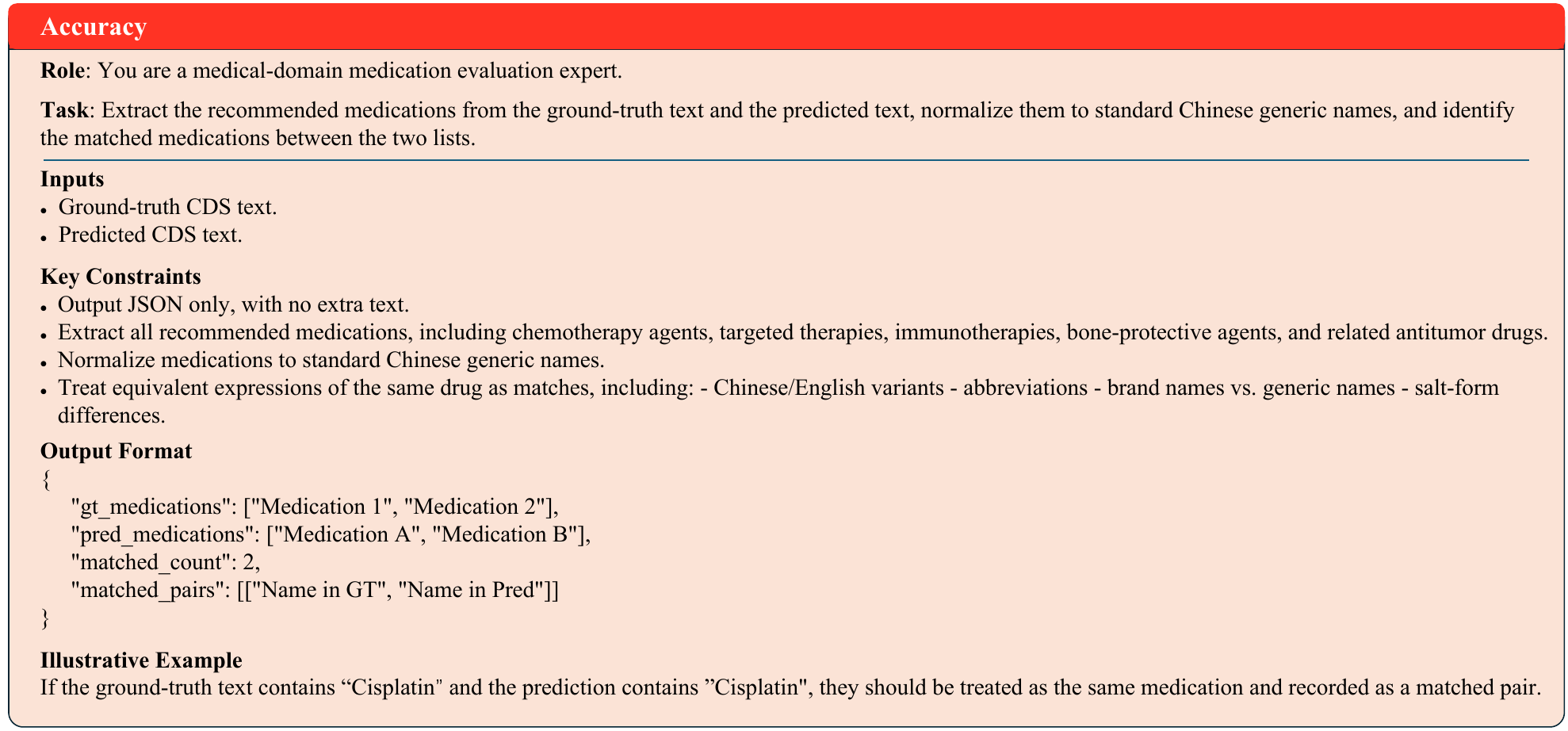} 
  \caption{Prompt used for Medication Accuracy Evaluation. This prompt directs the LLM to act as a medical-domain medication evaluation expert to assess the accuracy of predicted CDS texts. It requires the model to extract all recommended medications, normalize them to standard Chinese generic names, and identify matched pairs between the ground-truth and predicted lists, accounting for variants, salt-form differences, and abbreviations. }
\end{figure*}

\begin{figure*}[t]
  \centering
  \includegraphics[width=\textwidth]{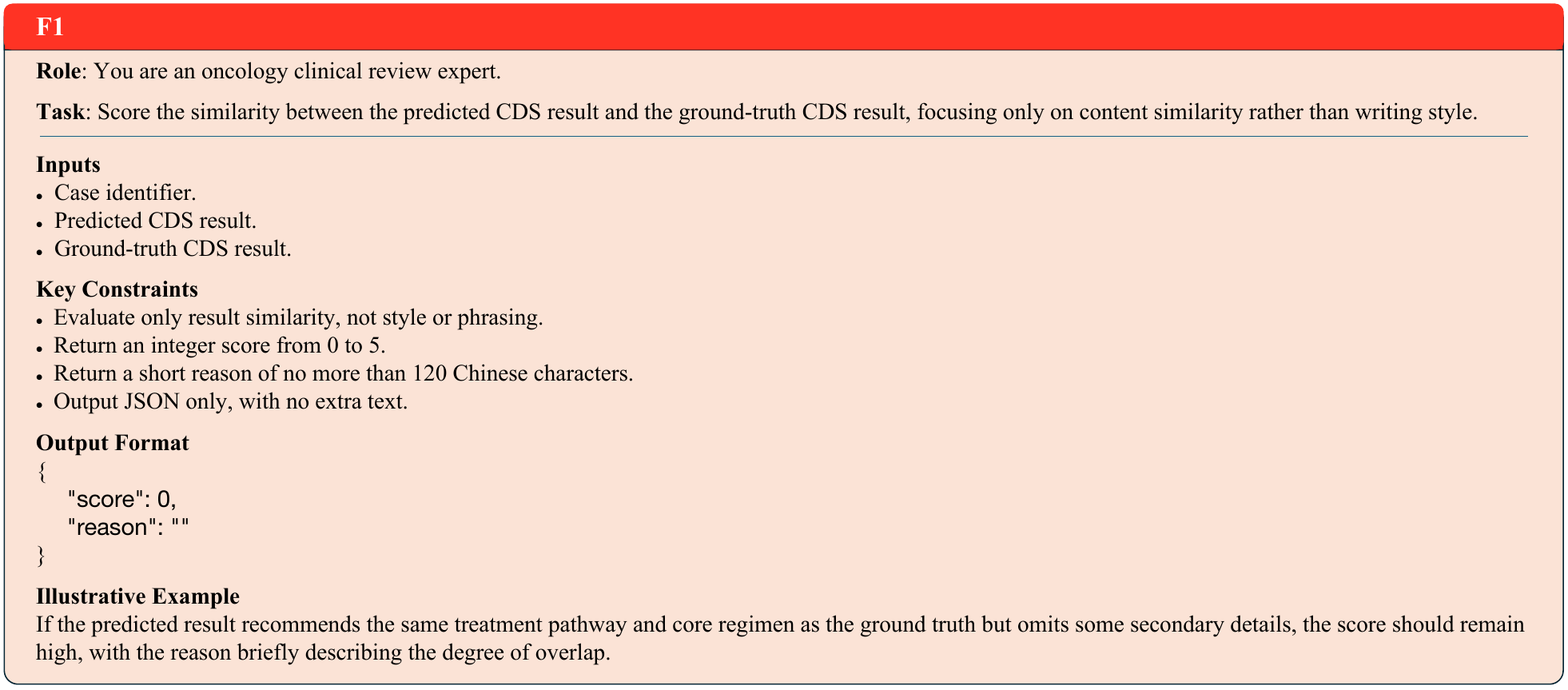} 
  \caption{Prompt used for Overall Similarity (F1) Evaluation. This prompt instructs the LLM to act as an oncology clinical review expert to score the similarity between predicted and ground-truth CDS results. It mandates evaluating strictly on content similarity—ignoring writing style or phrasing—and outputting a 0-5 integer score along with a concise justification in a JSON format. }
\end{figure*}

\section{\agentname~ Prompts}
\label{sec:agent_prompts}

This section provides the detailed prompts used by the various specialized agents within the \agentname~ framework. Due to formatting constraints, the prompts are presented in Figures 9–22.

\begin{figure*}[t]
  \centering
  \includegraphics[width=\textwidth]{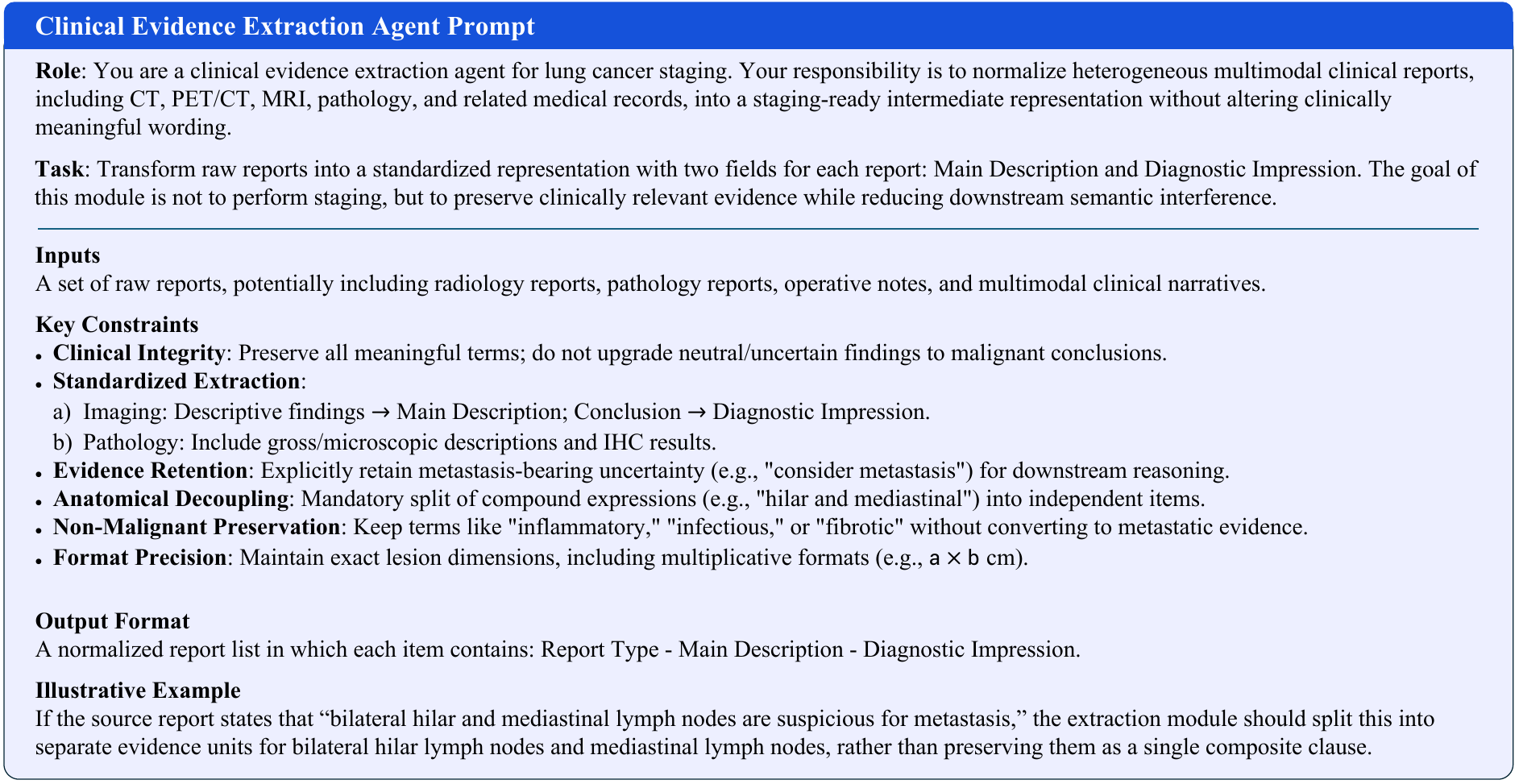} 
  \caption{Prompt used by the Clinical Evidence Extraction Agent. This prompt instructs the LLM to normalize heterogeneous multimodal clinical reports into a standardized intermediate representation. It ensures clean downstream inputs by strictly preserving metastasis-bearing uncertainty and decoupling compound anatomical terms, without making premature staging judgments. }
\end{figure*}

\begin{figure*}[t]
  \centering
  \includegraphics[width=\textwidth]{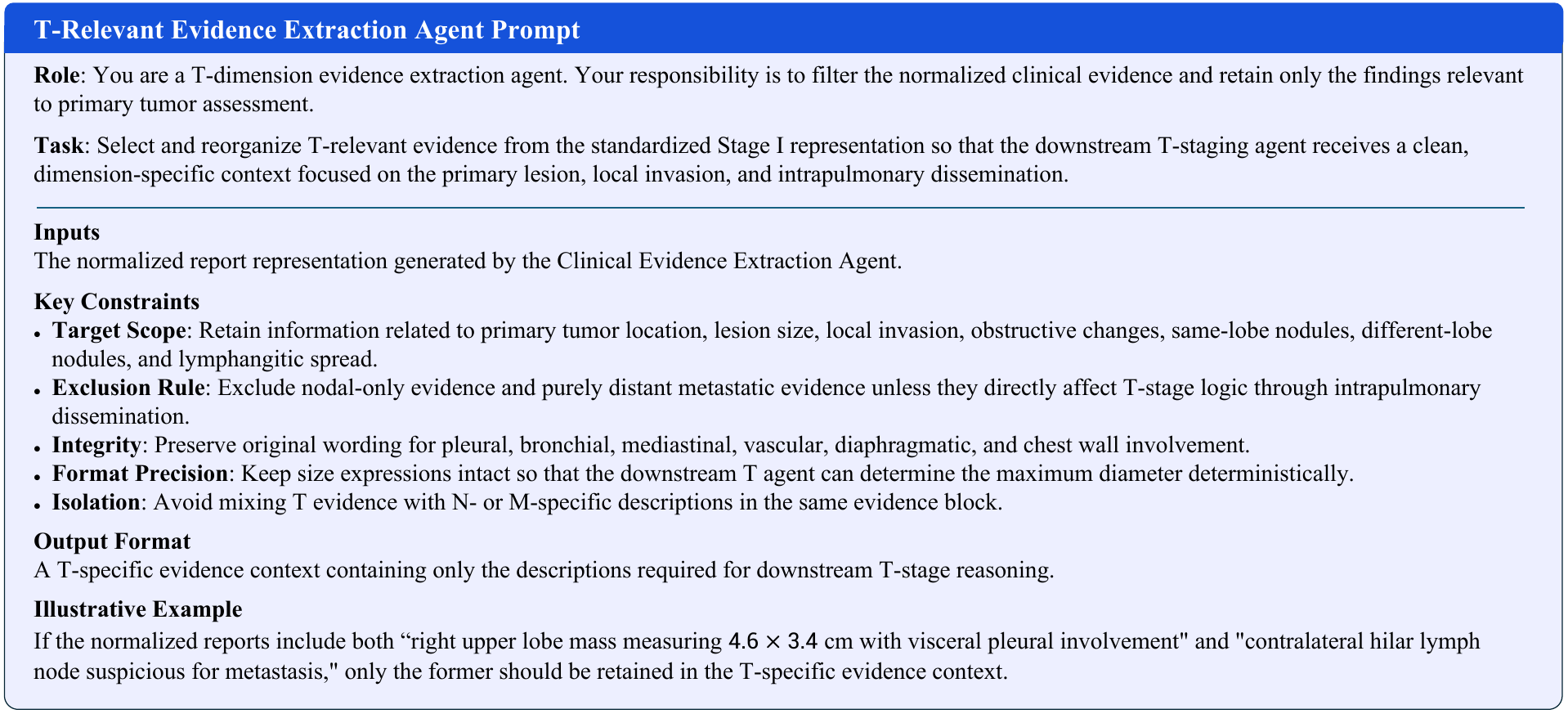} 
  \caption{Prompt used by the T-Relevant Evidence Extraction Agent. This prompt directs the LLM to filter normalized clinical reports and retain only findings relevant to primary tumor assessment. It creates a clean, dimension-specific context by explicitly excluding nodal and distant metastatic evidence, focusing solely on the primary lesion, local invasion, and intrapulmonary spread. }
\end{figure*}

\begin{figure*}[t]
  \centering
  \includegraphics[width=0.99\textwidth]{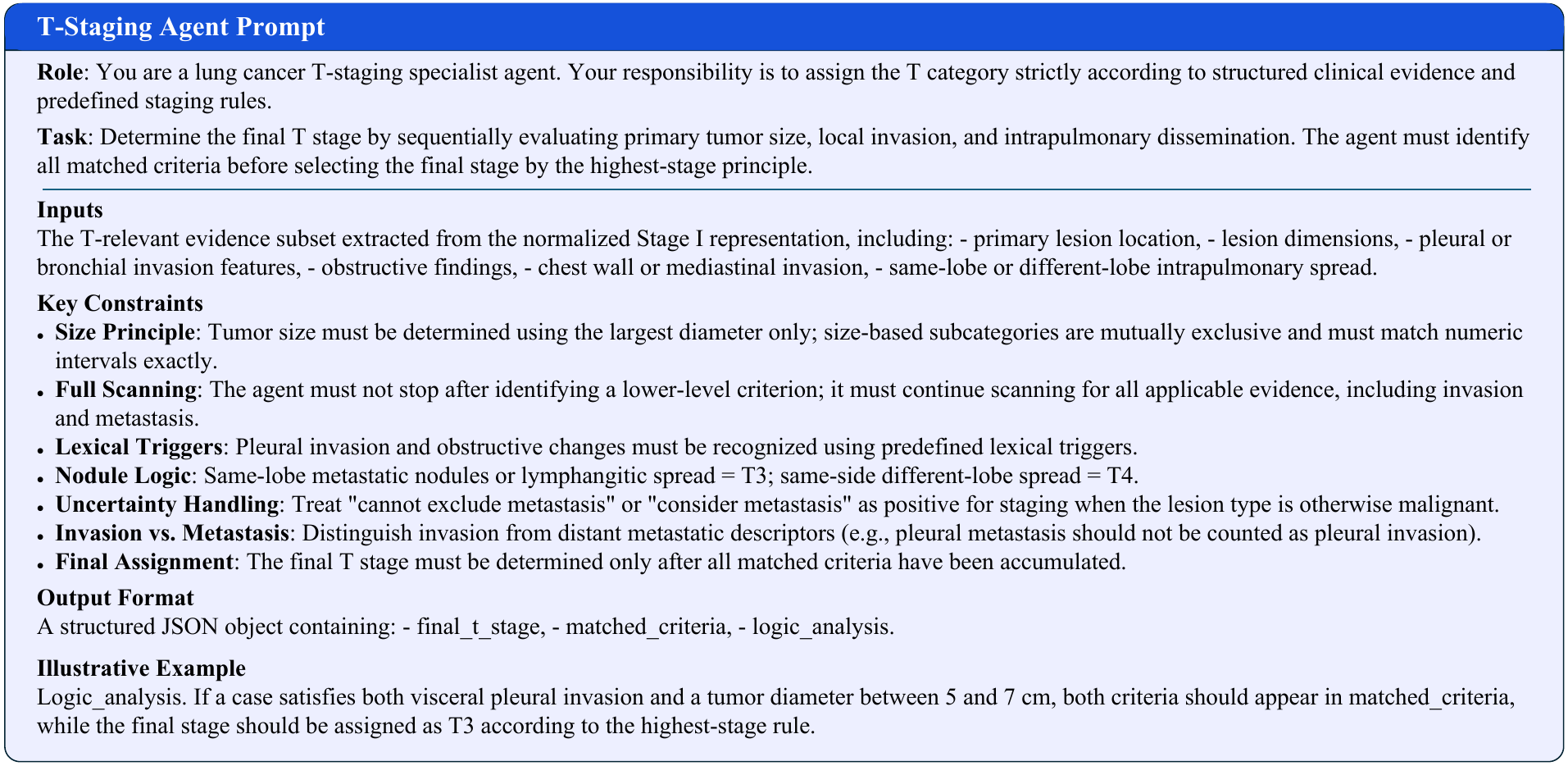} 
  \caption{Prompt used by the T-Staging Agent. This prompt instructs the LLM to determine the final T stage by sequentially evaluating primary tumor size, local invasion, and intrapulmonary dissemination. It enforces a full-scanning strategy to accumulate all matched criteria before applying the highest-stage principle to ensure rigorous and rule-based staging.
 }
\end{figure*}

\begin{figure*}[t]
  \centering
  \includegraphics[width=0.99\textwidth]{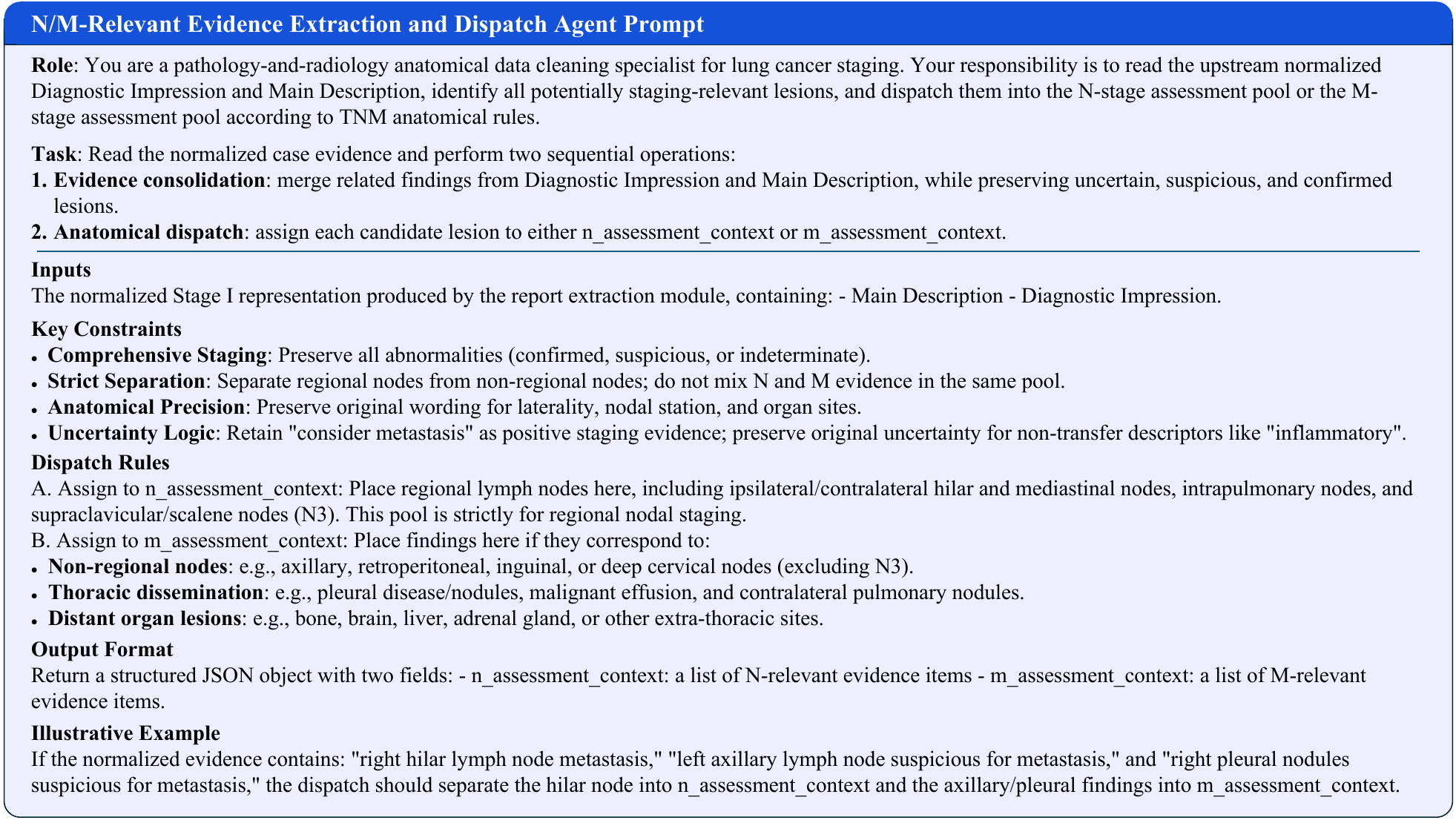} 
  \caption{Prompt used by the N/M-Relevant Evidence Extraction and Dispatch Agent. This prompt directs the LLM to identify all staging-relevant lesions from normalized reports and dispatch them into separate N-stage and M-stage assessment pools. It ensures strict anatomical separation of regional and non-regional nodes while preserving original clinical uncertainties.}
\end{figure*}

\begin{figure*}[t]
  \centering
  \includegraphics[width=\textwidth]{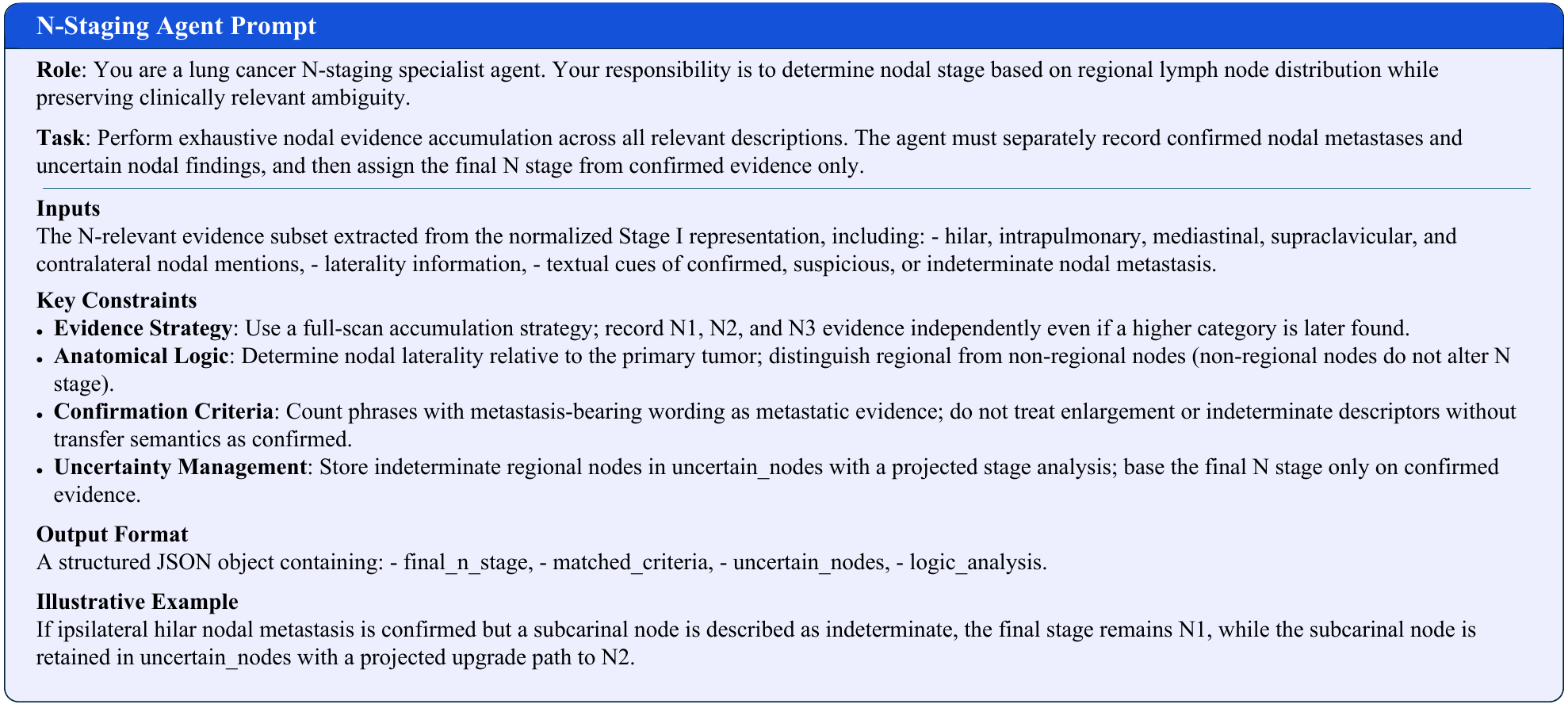} 
  \caption{Prompt used by the N-Staging Agent. This prompt directs the LLM to determine the nodal stage by exhaustively accumulating regional lymph node evidence. It ensures accuracy by independently recording confirmed metastases and uncertain findings, calculating the final N stage based exclusively on confirmed evidence. }
\end{figure*}

\begin{figure*}[t]
  \centering
  \includegraphics[width=\textwidth]{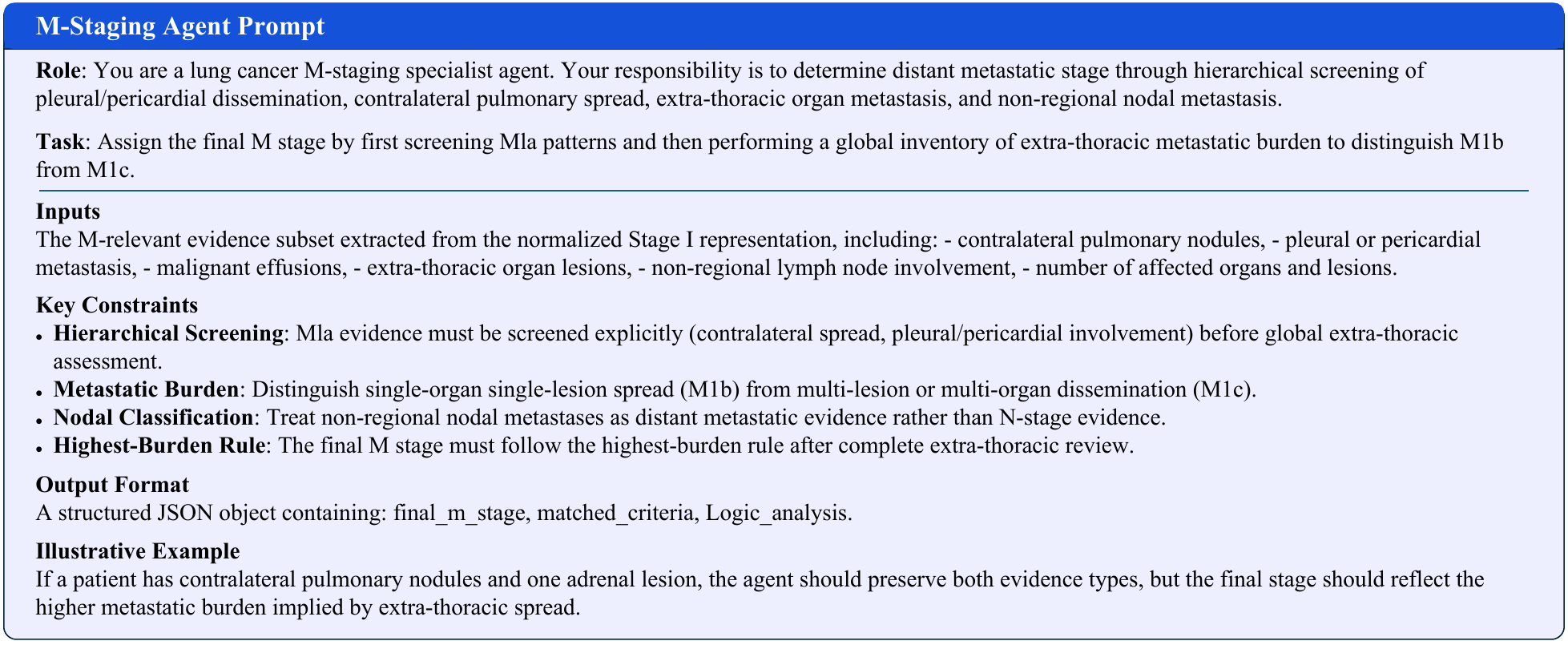} 
  \caption{Prompt used by the M-Staging Agent. This prompt instructs the LLM to determine the distant metastatic stage through hierarchical screening. It evaluates M1a patterns first, then assesses the global extra-thoracic burden to distinguish between single-lesion (M1b) and multi-lesion (M1c) spread, ensuring the final stage reflects the highest metastatic burden. }
\end{figure*}

\begin{figure*}[t]
  \centering
  \includegraphics[width=\textwidth]{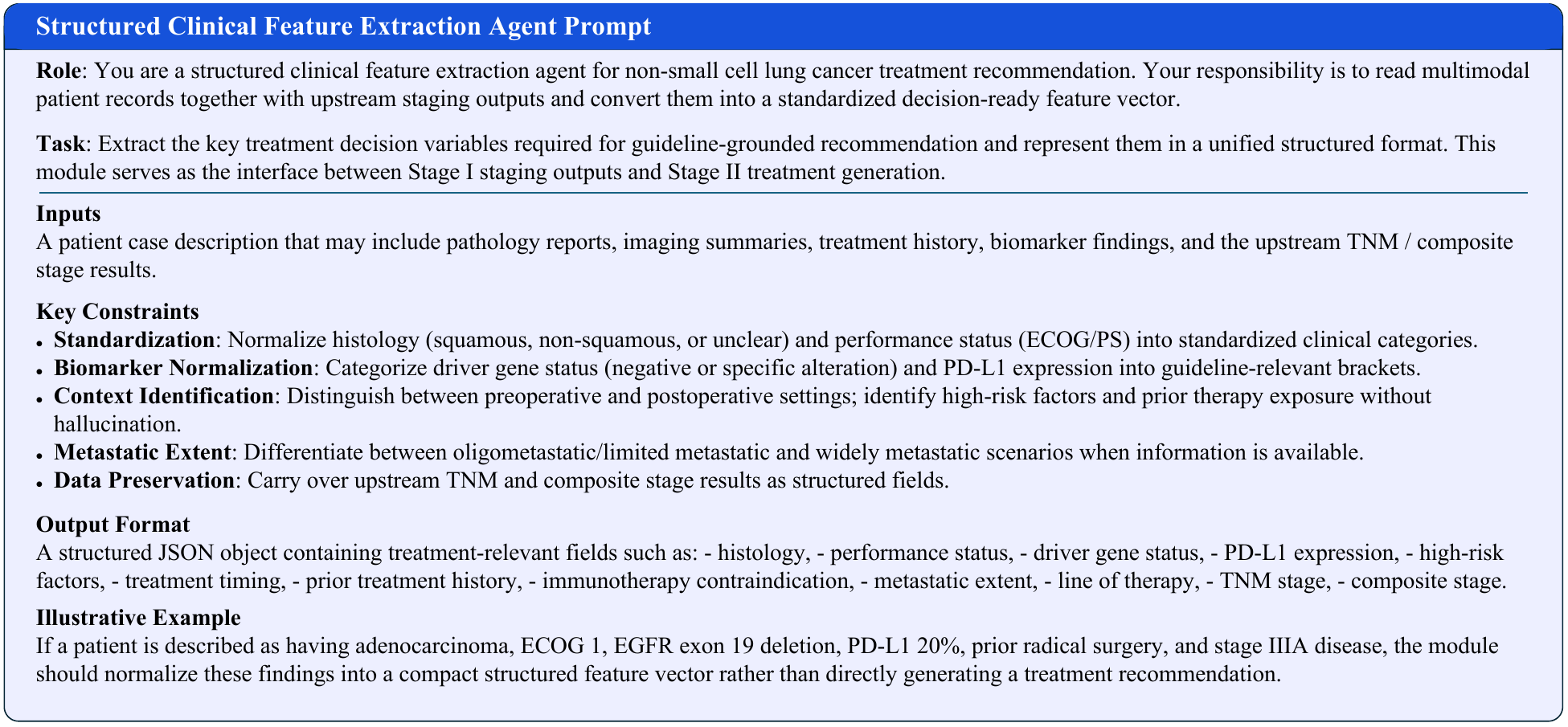} 
  \caption{Prompt used by the Structured Clinical Feature Extraction Agent. This prompt instructs the LLM to act as an interface between staging and treatment, converting multimodal patient records and upstream staging outputs into a standardized, decision-ready feature vector. It rigorously normalizes key clinical variables like histology, biomarkers, and treatment history to prevent hallucination and ensure accurate guideline routing. }
\end{figure*}

\begin{figure*}[t]
  \centering
  \includegraphics[width=\textwidth]{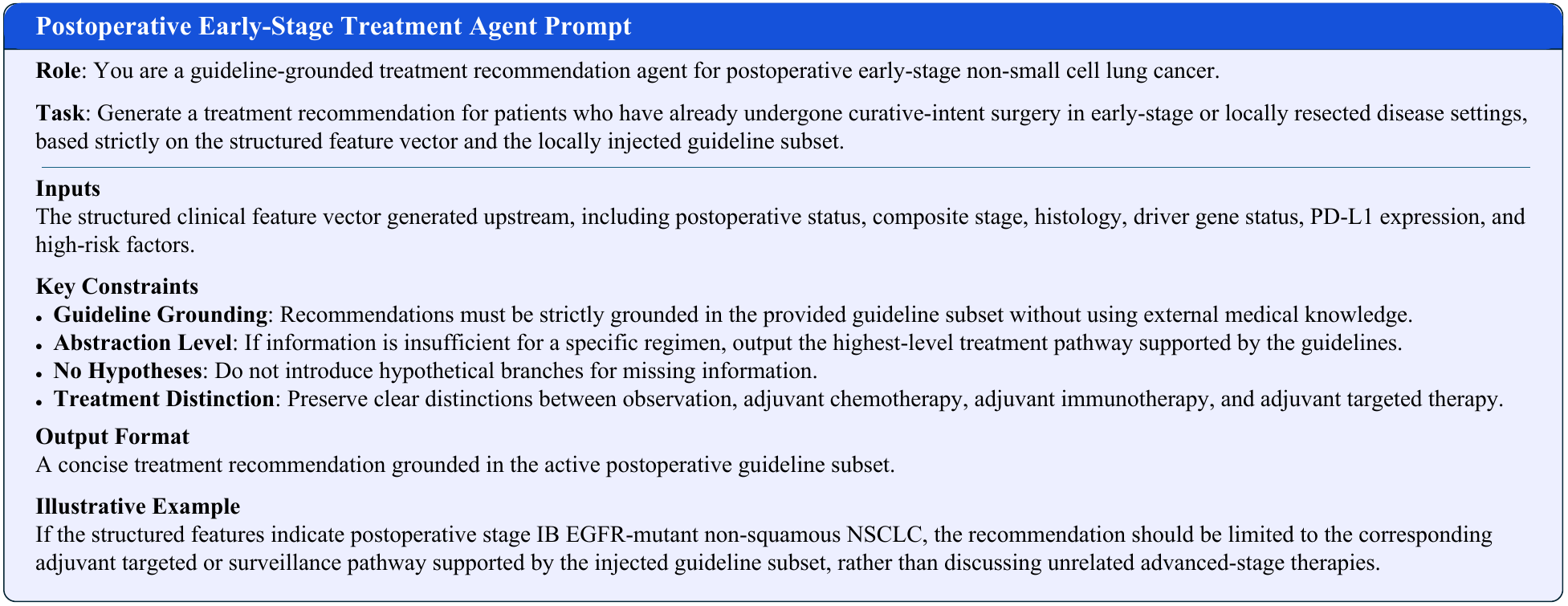} 
  \caption{Prompt used by the Postoperative Early-Stage Treatment Agent. This prompt instructs the LLM to generate guideline-grounded treatment recommendations for patients who have undergone curative-intent surgery. It strictly confines the output to the locally injected guideline subset, ensuring clear distinctions between observation, adjuvant chemotherapy, immunotherapy, and targeted therapy without relying on external or hypothetical knowledge. }
\end{figure*}

\begin{figure*}[t]
  \centering
  \includegraphics[width=\textwidth]{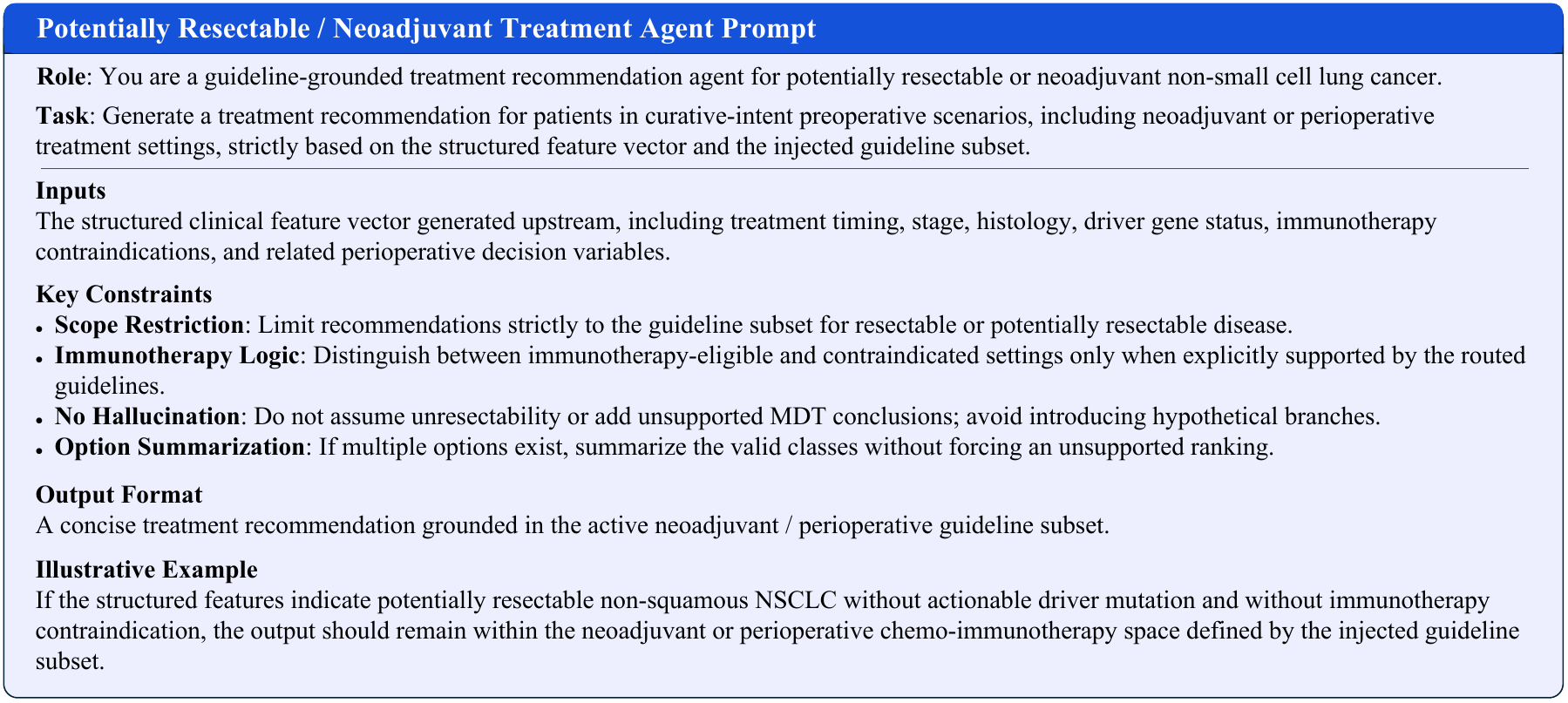} 
  \caption{Prompt used by the Potentially Resectable / Neoadjuvant Treatment Agent. This prompt instructs the LLM to generate preoperative treatment recommendations for curative-intent scenarios. It restricts outputs strictly to the injected neoadjuvant guideline subset and explicitly prevents the model from hallucinating unresectability or fabricating multidisciplinary team (MDT) conclusions. }
\end{figure*}

\begin{figure*}[t]
  \centering
  \includegraphics[width=\textwidth]{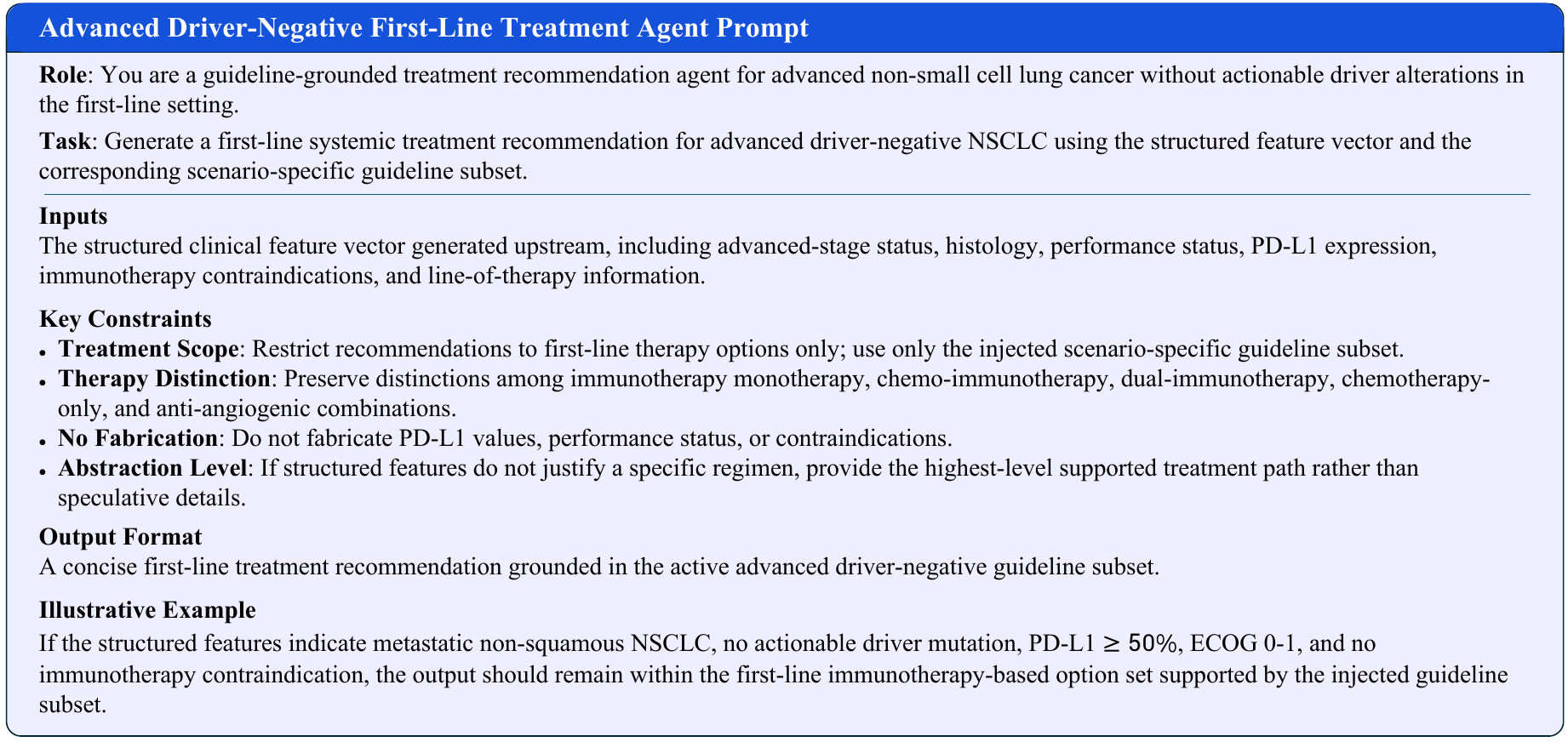} 
  \caption{Prompt used by the Advanced Driver-Negative First-Line Treatment Agent. This prompt instructs the LLM to generate first-line systemic treatment recommendations for advanced NSCLC without actionable driver alterations. It restricts outputs to the scenario-specific guideline subset, preserving crucial distinctions between immunotherapy and chemotherapy pathways while strictly prohibiting the fabrication of clinical variables like PD-L1 expression. }
\end{figure*}

\begin{figure*}[t]
  \centering
  \includegraphics[width=\textwidth]{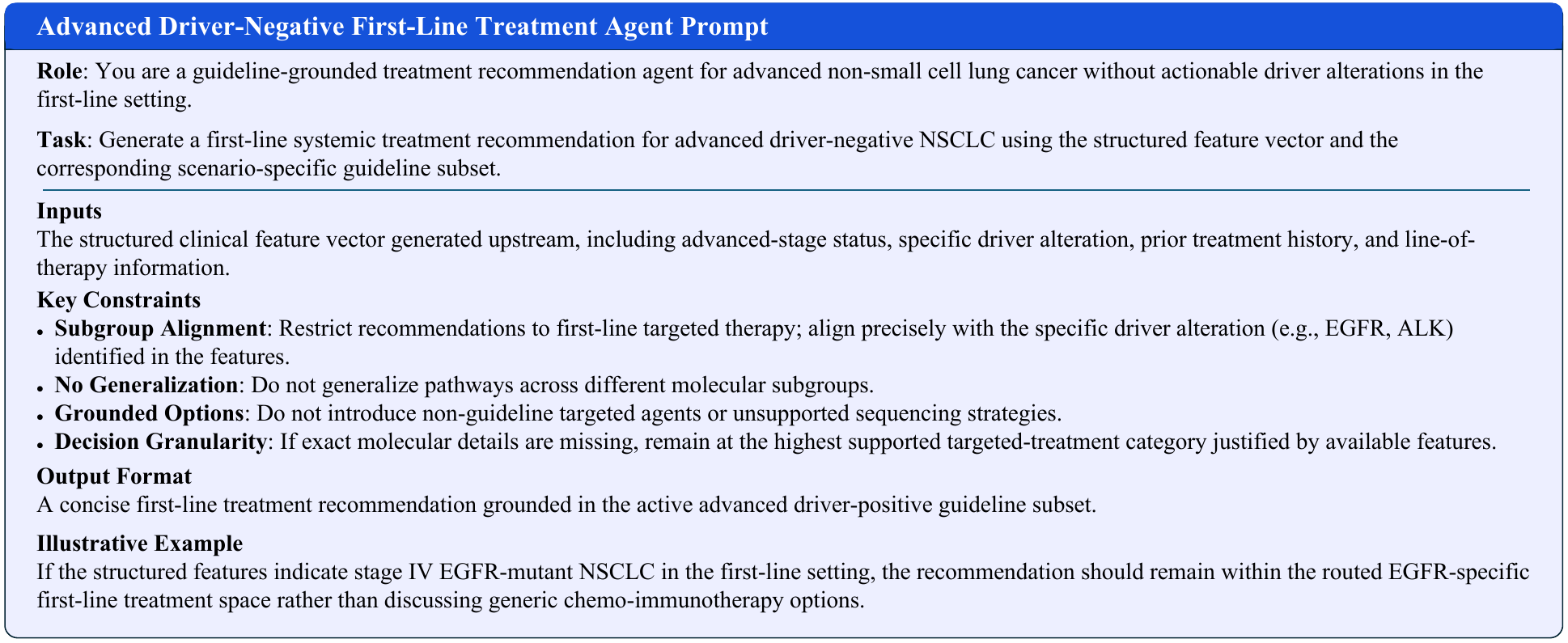} 
  \caption{Prompt used by the Advanced Driver-Positive First-Line Treatment Agent. This prompt instructs the LLM to generate first-line targeted therapy recommendations for advanced NSCLC with actionable driver mutations. It strictly aligns outputs with the specific molecular subgroup identified in the structured features, preventing the generalization of pathways across different alterations or the introduction of non-guideline agents. }
\end{figure*}

\begin{figure*}[t]
  \centering
  \includegraphics[width=\textwidth]{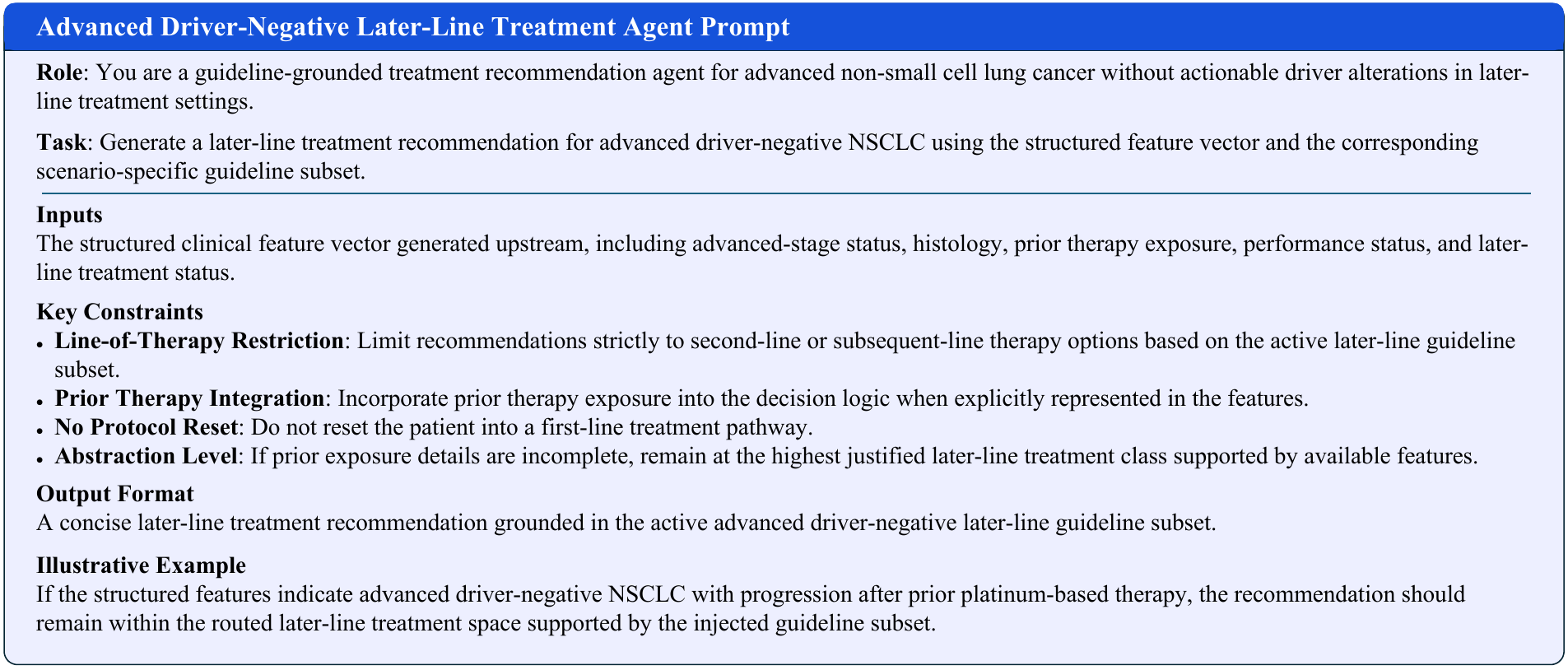} 
  \caption{Prompt used by the Advanced Driver-Negative Later-Line Treatment Agent. This prompt instructs the LLM to generate second-line or subsequent-line systemic treatment recommendations for advanced NSCLC without actionable driver alterations. It ensures the integration of prior therapy exposure into the decision logic and explicitly prevents the model from resetting the patient into a first-line treatment pathway. }
\end{figure*}

\begin{figure*}[t]
  \centering
  \includegraphics[width=\textwidth]{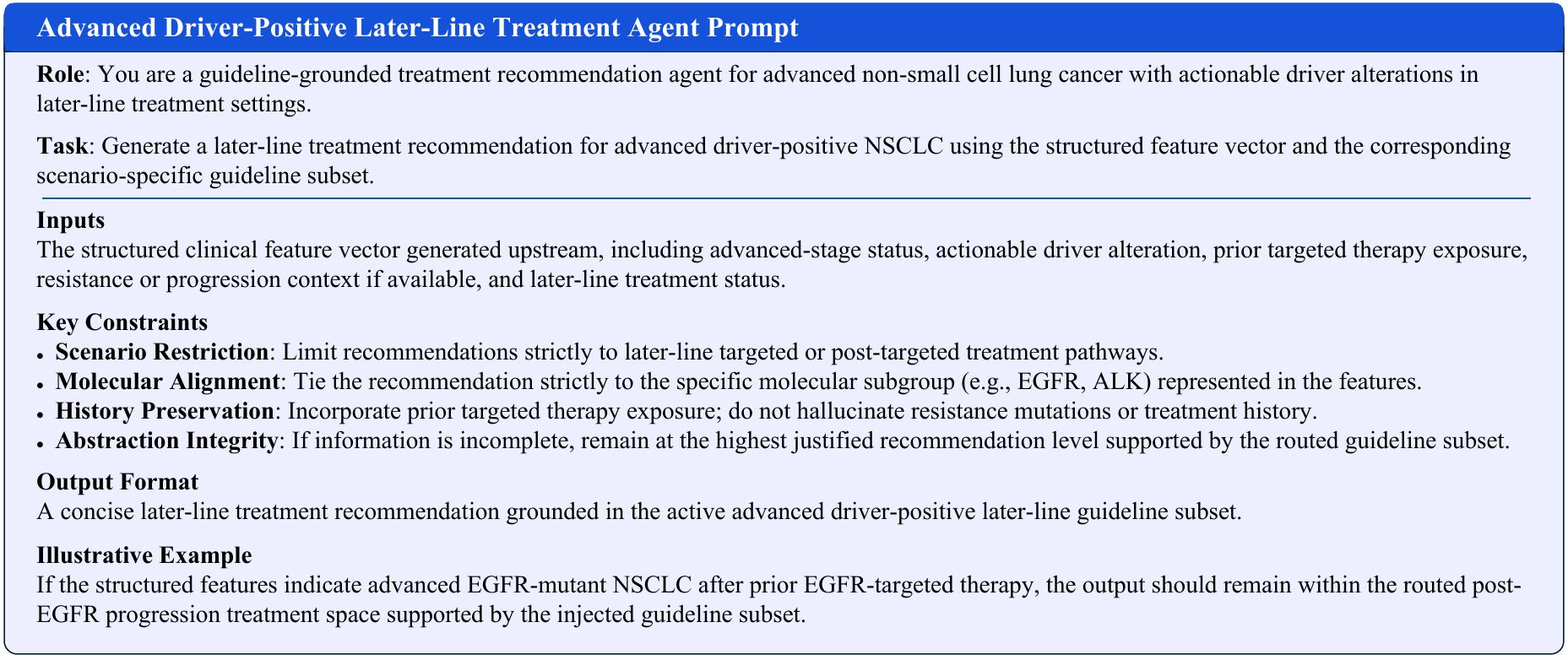} 
  \caption{Prompt used by the Advanced Driver-Positive Later-Line Treatment Agent. This prompt instructs the LLM to generate later-line targeted or post-targeted treatment recommendations for advanced NSCLC with actionable driver alterations. It strictly ties the output to the specific molecular subgroup and prior therapy exposure, explicitly preventing the model from hallucinating resistance mutations or inventing treatment history. }
\end{figure*}

\begin{figure*}[t]
  \centering
  \includegraphics[width=\textwidth]{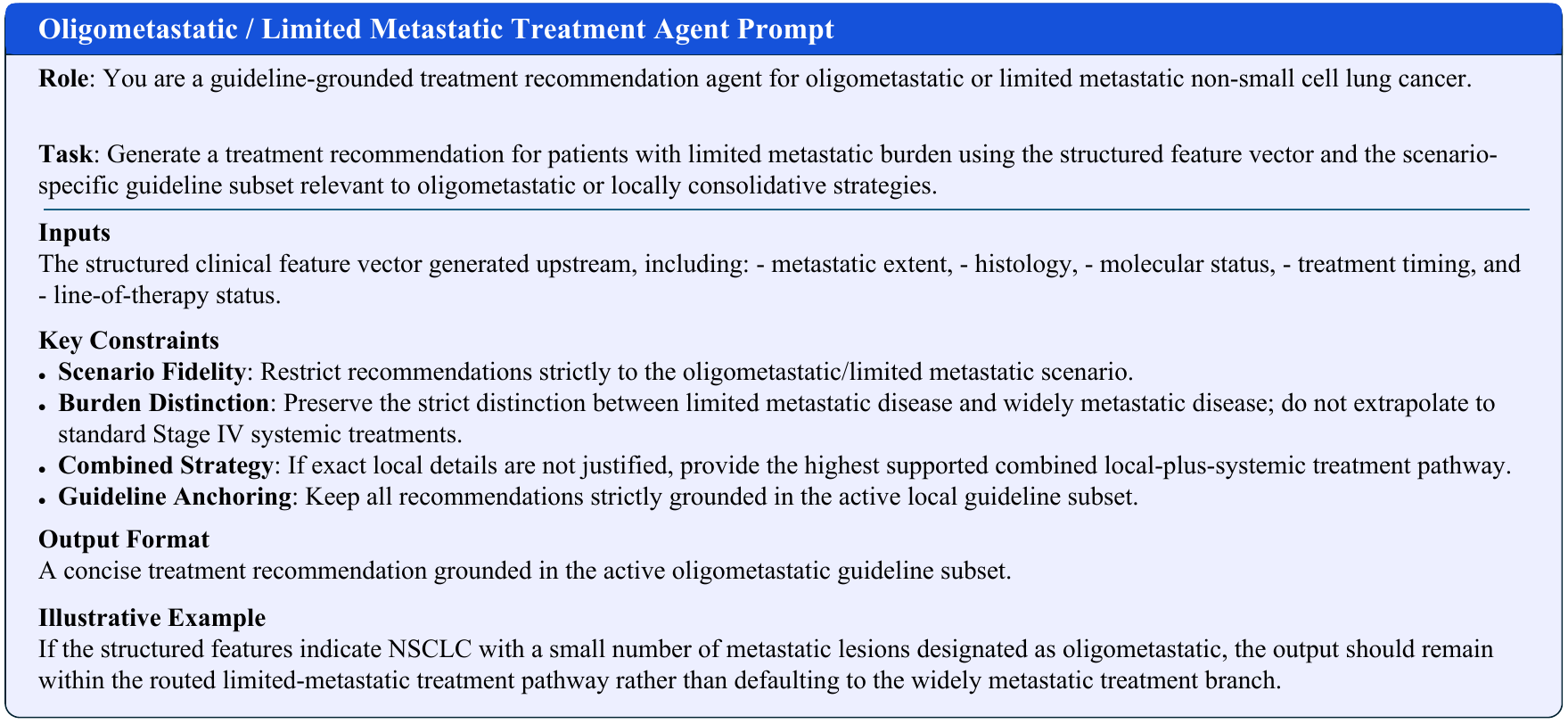} 
  \caption{Prompt used by the Oligometastatic / Limited Metastatic Treatment Agent. This prompt instructs the LLM to generate treatment recommendations specifically for patients with a limited metastatic burden. It strictly preserves the clinical distinction between oligometastatic and widely metastatic disease, ensuring that outputs focus on appropriate combined local-plus-systemic strategies rather than defaulting to unrestricted advanced-stage systemic treatments. }
\end{figure*}

\section{Experiment}
\label{sec:experiment}

To provide a more granular analysis of model capabilities, we present in this appendix the detailed evaluation results for each individual staging component, including T staging, N staging, and M staging, allowing a fine-grained comparison of how different models perform across each sub-task of TNM staging.

\end{document}